%% file: azzollini.tex
\documentclass{emulateapj}

\newcommand{\mstar}{$M_{\star}$\,}
\newcommand{\mb}{$M_{B}$\,}
\newcommand{\msun}{$M_{\sun}$\,}
\newcommand{\rbreak}{$R_{Br}$\,}
\newcommand{\mubreak}{$\mu_{Br}$\,}
\newcommand{\hone}{$h_{1}$\,}
\newcommand{\htwo}{$h_{2}$\,}
\newcommand{\mur}{$\mu$-r\,}
\newcommand{\magarcsq}{mag/$arcsec^{2}$\,}
\newcommand{\bb}{$B_{435}$\,}
\newcommand{\vv}{$V_{606}$\,}
\newcommand{\ii}{$i_{775}$\,}
\newcommand{\zz}{$z_{850}$\,}
\newcommand{\mbrange}{-19.5$>M_{B}>$-21 mag\,}
\newcommand{\mstarrange}{$5\cdot10^{9}<M_{\star}<5\cdot10^{10}$ \msun\,}

\slugcomment{}

\shorttitle{Cosmic Evolution of Stellar Disk Truncations}
\shortauthors{Azzollini, Trujillo \& Beckman}

\begin{document}


\title{Cosmic Evolution of Stellar Disk Truncations: \\
From \MakeLowercase{z}$\sim$1 to the Local Universe}


\author{R. Azzollini, I. Trujillo and J. E. Beckman\altaffilmark{1}}
\affil{Instituto de Astrof\'isica de Canarias, C/V\'ia L\'actea s/n, 38205 La Laguna, S/C de Tenerife, Spain}
\email{ruyman@iac.es, trujillo@iac.es, jeb@iac.es}


\altaffiltext{1}{Consejo Superior de Investigaciones Cient\'ificas, Spain}


\begin{abstract}
We have conducted the largest systematic search so far for stellar disk truncations in disk-like galaxies at intermediate redshift ($z$$<$1.1), using the Great Observatories Origins Deep Survey South (GOODS-S) data from the \emph{Hubble Space Telescope} - ACS. Focusing on Type II galaxies (i.e. downbending profiles) we explore whether the position of the break in the rest-frame $B$-band radial surface brightness profile (a direct estimator of the extent of the disk where most of the massive star formation is taking place), evolves with time. The number of galaxies under analysis (238 of a total of 505) is an order of magnitude larger than in previous studies. For the first time, we probe the evolution of the break radius for a given stellar mass (a parameter well suited to address evolutionary studies). Our results suggest that, for a given stellar mass, the radial position of the break has increased with cosmic time by a factor 1.3$\pm$0.1 between $z$$\sim$1 and $z$$\sim$0. This is in agreement with a moderate inside-out growth of the disk galaxies in the last $\sim$ 8 Gyr. In the same period of time, the surface brightness level in the rest-frame $B$-band at which the break takes place has increased by 3.3$\pm$0.2 \magarcsq (a decrease in brightness by a factor of 20.9$\pm$4.2). We have explored the distribution of the scale lengths of the disks in the region inside the break, and how this parameter relates to the break radius. We also present results of the statistical analysis of profiles of artificial galaxies, to assess the reliability of our results.
\end{abstract}





\section{Introduction}\label{sec1}

Early studies of the disks of spiral galaxies \citep{Patterson40, deVaucouleurs59,Freeman70} showed that this component generally follows an exponential radial surface-brightness profile, with a certain scale length, usually taken as the characteristic size of the disk. \citet{Freeman70} pointed out, though, that not all disks follow this simple exponential law.

Another repeatedly reported feature of disks is that of a truncation of the stellar population at large radii, typically 2-4 exponential scale lengths \citep[see e.g. the review by][]{Pohlen04}. \Citet{Kruit79} and \citet{KruitSearle81a,KruitSearle81b} first drew attention to this phenomenon, which they inferred primarily from the major axis profiles of edge-on, late-type spirals. Though the term ``truncation'' is used, not even in the original studies was a complete absence of stars beyond the truncation radius suggested. In fact, \citet{Pohlen02} showed that the truncation actually adopts the form of a fairly sharp change in slope, from the shallow exponential of the main disk to a steeper exponential at larger radii \citep[see also][]{deGrijs01}. \citet{Erwin05} denoted galaxies with this feature as Type II objects, generalizing a classification scheme initially proposed by \citet{Freeman70}. 

Though the truncation phenomenon appears to be very widespread \citep[see for example][hereinafter cited as PT06]{PT06}, there are many cases in which this just does not appear to happen even at extremely faint surface brightness levels. To give an example, \citet{Bland-Hawthorn05} found (using star counts) a galaxy (NGC 300) for which the single exponential decline simply continues down to $\sim$ 10 radial scale lengths ($\sim$ 30.5 \magarcsq in $r'$ band). Together with earlier measurements by e.g. \citet{BartonThompson97} or \citet{Weiner01} (using surface photometry) this provides evidence that indeed there are prototypical exponential disks \citep[Type I objects, following nomenclature in][]{Freeman70}. More strikingly, perhaps, there even exists evidence for a third type of profile presented by \citet{Erwin05,Erwin07,Erwin08} for early-type disks, by \citet{HunterElmegreen06} and PT06 for late-type disks ; and for extreme late-type spirals by \citet{MatthewsGallagher97}. In this class, named ``Type III'' (also ``antitruncations'') by \citet{Erwin05}, the inner profile is a relatively steep exponential, which gives way to a shallower profile, beyond the break radius. This profile is thus something like the ``inverse'' of a Type II profile, bending ``up'' instead of ``down'' beyond the break radius (``upbending'' profile).

In this paper we will concentrate on the Type II galaxies. Several possible break-forming mechanisms have been investigated to explain their truncations. There have been ideas based on maximum angular momentum distribution: \citet{Kruit87} proposed that angular momentum conservation in a collapsing, uniformly rotating cloud naturally gives rise to disk breaks at roughly 4.5 scale radii. \Citet{Bosch01} suggested that the breaks are due to angular momentum cut-offs of the cooled gas. On the other hand, breaks have also been attributed to a threshold for star formation (SF), due to changes in the gas density \citep{Kennicutt89}, or to an absence of equilibrium in the cool Interstellar Medium phase \citep{ElmegreenParravano94,Schaye04}. Following this, and using a semi-analytic model, \citet{ElmegreenHunter06} demonstrated that a double-exponential profile may arise from a multi-component star formation prescription. The above two ``simple'' scenarios (angular momentum vs. star formation threshold), however, are challenged by observational results. It is becoming more and more clear that galaxies have a significant density of stars beyond the break radius. In addition, the existence of extended UV disks \citep{Thilker05,Gildepaz05} and the lack of a clear correlation between H$\alpha$ cut-offs and optical disk breaks \citep{Pohlen04,HunterElmegreen06} further complicates the picture. Even though a sharp star formation or angular momentum cut-off may explain a disk truncation, it does not provide a compelling explanation for extended outer exponential components. More elaborate models such as that by \citet{Debattista06} demonstrated that the redistribution of angular momentum by spirals during bar formation also produces realistic breaks, using collisionless N-body simulations. In a further elaboration of this idea, \citet{Roskar08} have performed high resolution simulations of the formation of a galaxy embedded in a dark matter halo \citep[see also][]{Bournaud07,Foyle08}. They are able to reproduce Type II profiles, which they claim to be in good agreement with observations. In their model, breaks are the result of the interplay between a radial star formation cut-off and redistribution of stellar mass by secular processes.

Independently of the formation mechanism of the truncation, it seems reasonable to assume that the structural properties of the faintest regions of galactic disks must be intimately linked to the processes involved in the growing and shaping of galaxies. These outer edges are easily affected by interactions with other galaxies and, consequently, their characteristics must be closely connected with the evolutionary paths followed by the galaxies \citep{PT06,Erwin07}. Together with their stellar halos, the study of the outer edges allows the exploration of the ``fossil'' evidence imprinted by the galaxy formation process \citep{deJong07}.

Furthermore, addressing the question of how the radial truncation evolves with z is strongly linked to our understanding of how the galactic disks grow and where star formation takes place. \citet{Perez04} showed that it is possible to detect stellar truncations even out to $z\sim1$. Using the radial position of the truncation (hereafter \rbreak) as a direct estimator of the size of the stellar disk, \citet{TP05} inferred a moderate ($\sim $25\%) inside-out growth of disk galaxies since $z\sim1$. An important point, however was missing in the previous analyses: the evolution with redshift of the radial position of the break at a given stellar mass. The stellar mass is a much better parameter to explore the growth of galaxies, since the luminosity evolution of the stellar populations can mimic a size evolution \citep{Trujillo04,Trujillo06}. Addressing this point here, one aim of the present work is to probe whether the galaxies are growing from the inside outwards, with star formation propagating radially outward with time. Another point to mention is that while \citet{TP05}, worked with imaging data from the Ultra Deep Field \citep{Beckwith06}, we make use of GOODS-HST/ACS data \citep{Giavalisco04}. This field provides a much wider sky coverage, and so increases the size of the sample under study by roughly an order of magnitude, though at the cost of using images with somewhat lower signal to noise ratio. We take care to tackle the implications of this caveat in the paper as well.

In this work (to which we will refer hereinafter as ATB08) we also broach other issues relative to the detailed characterization of the surface brightness profiles (\mur profiles). We present results on the distribution of the $B$-band rest-frame surface brightness at the position of the break (hereafter \mubreak). We discuss the distribution of the ratio \rbreak / \hone, where \hone is the scale length of the disk in the region inner to the break. Also included are results on the relation between \hone and absolute magnitude and stellar mass. Finally, for the first time in this kind of studies, we make extended analysis of the profiles of simulated galaxies, in an attempt to better limit the accuracy of the reported figures.

The outline of this paper is as follows. In section \ref{sec2} we describe the galaxy sample and give a brief description of the imaging data we use. In Section \ref{sec3} we describe how we obtained the radial surface-brightness profiles of the galaxies, and the methodology of the analysis we apply for their characterization. Section \ref{sec4} is focused on the completeness of the samples under study and assessing the accuracy level of the parameters retrieved. Section \ref{sec5} presents results on classification and characterization of the \mur profiles, and more extensive analysis of how the Break Radii of Type II disks relate to various other properties of the galaxies at different redshifts. In section \ref{sec6} we summarize our results and conclusions. Finally we include an Appendix on the description and analysis of results from an exercise of classification of profiles performed on a sample of artificial model \emph{galaxies}. Throughout, we assume a flat $\Lambda$-dominated cosmology ($\Omega_{M}$ = 0.30, $\Omega_{\Lambda}$=0.70, and $H_{0}$=70 km $s^{-1}$ $Mpc^{-1}$).

\section{DATA \& SAMPLE SELECTION}\label{sec2}





We have worked with objects present in imaging data from HST-ACS observations of the GOODS-South field \citep{Giavalisco04}, which subtends an area in the sky of roughly 170 square arcmin. These data are publicly available\footnote{\url{http://www.stsci.edu/ftp/science/goods/}}, and consist of imaging in the $F435W$, $F606W$, $F775W$ and $F850LP$ HST pass-bands, hereafter referred to as \bb, \vv, \ii and \zz, and whose total exposure times are approximately 7200, 3040, 3040, and 6280 seconds, respectively. The images have an angular scale of 0.03''/pixel and the FWHM of the PSF in \zz measures 0.09''.

To make the selection of our sample we have profited from the fact that the GOODS-South field is within the \emph{Galaxy Evolution from Morphologies and SEDs} imaging survey \citep[GEMS;][]{Rix04}. In the redshift range 0.1$\leq$$z$$\leq$1.1, GEMS provides morphologies and structural parameters for nearly 10,000 galaxies \citep{Barden05, McIntosh05}. For many of these objects there also exist photometric redshift and luminosity estimates, and SEDs from COMBO-17 \citep[\emph{Classifying Objects by Medium-Band Observation in 17 filters};][]{Wolf01, Wolf03}. The COMBO-17 team made this information publicly available through a catalog with precise redshift estimates (with errors $\delta$$z$/(1+$z$)$\sim$0.02) for approximately 9000 galaxies down to $m_{R}<24$ \citep{Wolf04}. In the same data release were included rest-frame absolute magnitudes and colors (accurate to $\sim$ 0.1 mag). We have also used the stellar mass estimates published in \citet{Barden05}, which are taken from \citet{Borch04}, and are deduced from the COMBO-17 photometric data.

\citet{Barden05} conducted the morphological analysis of the late-type galaxies in the GEMS field by fitting S\'ersic $r^{1/n}$ \citep{Sersic68} profiles to the surface brightness distributions. \citet{Ravindranath04} showed that using the S\'ersic index $n$ as the criterion, it is feasible to distinguish between late- and early-type galaxies at intermediate redshifts. Late-types (Sab-Sdm) are defined as having $n$$<$2-2.5. Moreover, the morphological analysis conducted by Barden et al. provides the information about the inclination of the galaxies. This is particularly important, since we want to study the truncation of the stellar disks in objects with low inclination. The edge-on view facilitates the discovery of truncations but introduces severe quantitative problems caused by the effects of dust and line-of-sight integration which we want to avoid \citep{Pohlen02}.

We selected objects from the Barden et al. sample within the subsequent ranges of parameters: S\'ersic index $n$$\leq$2.5 to isolate disk-dominated galaxies \citep{Barden05, Shen03, Ravindranath04}; axial ratio $q$ $>$ 0.5 to select objects with inclination $<$ 60$\arcdeg$; and \mb $<$ -18.5 magnitudes, as in \citet{TP05}. Moreover, only objects with $z$$<$1.1 were selected in order to maintain our analysis in the optical rest-frame bands. Finally the resulting sample was matched to a photometric catalog derived by ourselves from the GOODS-South HST-ACS data (GOODS data hereafter), using SExtractor\footnote{This catalog was obtained by detecting sources in the \zz band which had at least 16 contiguous pixels (0.014 $arcsec^{2}$) at an isophotal level of 0.6 sky $\sigma$ (25.35 \magarcsq) or higher. These photometry parameters are the same employed by the GOODS Teams for producing their catalogs. Nonetheless, we wanted to have segmentation maps for each object, as produced by SExtractor, but the GOODS team did not release these files, and so we had to do our own, though equivalent, photometry} \citep[][]{BA96}. The resulting sample contains 505 objects.

To analyse the surface brightness profiles of our galaxies in a similar rest-frame band within the explored redshift range (0.1$<$$z$$<$1.1), we have extracted the profiles in the following bands: $V_{606}$ band for galaxies with 0.1$<$$z$$\leq$0.5, $i_{775}$ band for 0.5$<$$z$$\leq$0.8, and $z_{850}$ band for 0.8$<$$z$$\leq$1.1. This allows us to explore the surface brightness distribution in a wavelength close to the $B$-band rest-frame.

\section{RADIAL PROFILE ANALYSIS}\label{sec3}

The surface brightness (\mur) profiles were extracted using photometry on quasi-isophotal elliptical apertures. The intensities were estimated as the median of the flux in the area between elliptical apertures of increasing semi major axis length (hereafter we will dub these lengths as ``radii''). The ellipticity and position angle of the apertures were fixed to those retrieved by SExtractor for the whole distribution of light of the object. The center of the apertures was also fixed: in a first iteration, the first moments (in `$x$' and `$y$') of the brightness distribution of flux of the object were used for this purpose. After visual inspection of the resulting profile and the image of the object, the center was refined, if needed, with help of the task ``imexam'' from iraf\footnote{\url{http://iraf.noao.edu}} to match what was visually estimated, in each case, as the dynamical center of the object. In a regular disk galaxy, as are those under study, this center coincides with the central bulge or nucleus. The radii of the annular apertures was linearly increased at constant steps of 1 pixel (0.03''), up to a radius which is a 50\% larger than the radius of a circle with the same area as the isophotal area of the object in the $z_{850}$ band, as given by SExtractor. The isophotal area is that covered by the set of connected pixels with intensities above the detection threshold which constitute a detection (in our case 25.4 \magarcsq in \zz). For objects with a more or less regular morphology, as are those selected, the median intensities in these annulii are a good approximation to isophotal intensities. The error in the intensity, $\delta$I, is given by the $\sigma$ of the distribution of fluxes inside the annulus, divided by the square root of the corresponding pixel area. The intensities (``$I$'') were transformed to surface brightnesses ($\mu$), expressed in \magarcsq in the AB system, using the magnitude zero points posted on the GOODS-HST/ACS website for each filter and the angular scale per pixel, through the expression $\mu = zero - 2.5 \cdot log(I/scale^{2})$. For the errors in $\mu$, $\delta\mu$, the formula used was $\delta\mu$ = 2.5 $\cdot$ log( 1 + $\delta$$I$/$I$). In producing these profiles, SExtractor, DS9\footnote{\url{http://hea-www.harvard.edu/RD/ds9/}} and Iraf software packages were used, ``glued'' together by a script written by ourselves in ``Python''\footnote{\url{http://www.python.org}} language.
                                                   
We produced radial profiles, and characterized them in terms of the properties of the exponential laws which best fit them. For each object, the image and profile obtained are visually inspected. If the object is affected by artifacts (which mostly occurs near the edges of the area of the GOODS-HST-ACS field), or the object is suspected of being a merger, or has an irregular morphology, or the object is between two tiles of the survey, or part of the object lies outside the field, the object is rejected for further analysis. From the initial 505 objects, 70 were rejected for these reasons (14\%).

For those objects which are suitable for analysis, the profile was tested for the existence of breaks of any kind, trying the 3 methods we describe next and comparing results, and also by visually judging the apparent morphology of the galaxy. If no break is apparent (Type I), the profile is fitted to a single exponential. If there is a break, it is characterized by using one of the two following methods. The first, which we call the ``Intersection method'', was the most commonly applied. It was used in 84\% of the 289 objects which show a ``break'' (i.e. Types II and III). In this method two exponentials were independently fitted to non overlapping sections of the profile. The point of intersection of the two lines gives the Break Radius, \rbreak, while the surface brightness on the profile, linearly interpolated at that radius, is the surface brightness at the break, \mubreak. The radial scales of the exponentials (\hone and \htwo) are used to distinguish between Types II (\hone $>$ \htwo) and III (\hone $<$ \htwo).

\begin{figure}
\epsscale{1}
\plotone{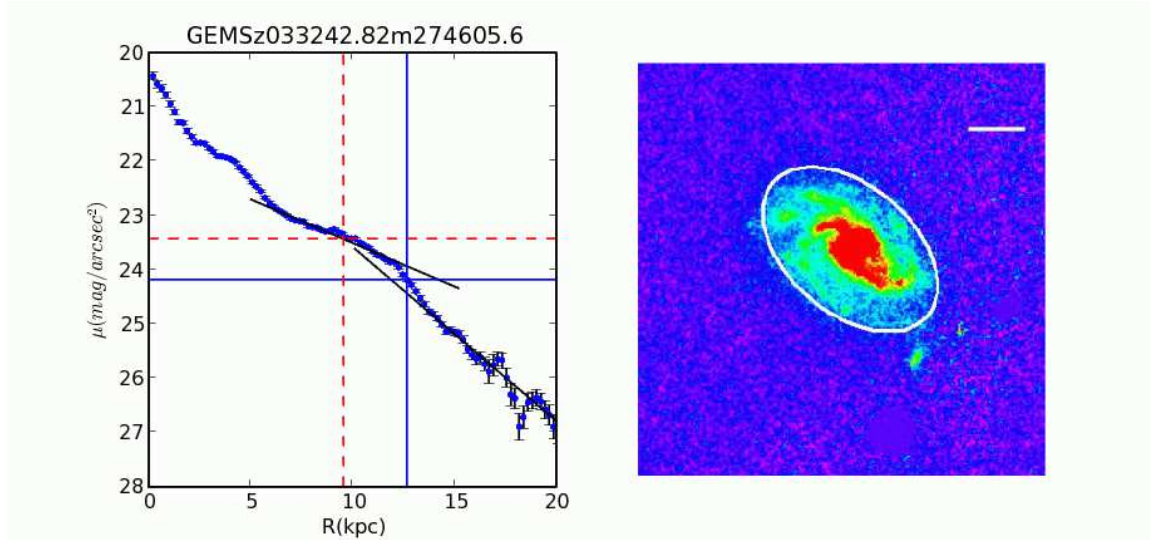}
\caption{Example of the kind of break we define as ``Jump Truncation'', in which a minor discontinuity in intensity appears between the ``inner'' and ``outer'' exponential regions of the profile. In the left panel the discontinuous lines mark the position of the break according to the ``Intersection Method'', while the continuous lines give the position given by the ``Equal Deviation'' method, specifically devised for these kind of profiles (16\% of the objects which show a break). The ellipse on the image marks the position of the break, while the white line in the upper right corner is 1 arcsec long in the image.\label{figJump}}
\end{figure}

In 47 cases (16\% of the objects which show a break), a slightly more involved process is required to measure the position and brightness of the break. We have found some objects which show a change in slope of some Type (mostly II, but also III), in which there is a smaller zone between the 2 exponential regions with a different slope from the other two more extended zones. An example of this phenomenology is shown in Fig.~\ref{figJump}. If the length of this intermediate zone were increased, and if its slope were constant, it would be classified as a ``mixed'' Type \citep[II+III or III+II,][PT06]{Erwin05}, but we refer here to those cases in which the radial extent of this intermediate zone is so small that it seems to be showing a specific morphology. Measuring the position of the break in these (from now on in this paper) termed ``Jump Truncations'', cannot be done with the ``Intersection method''. In these cases, the intersection point does not match the point at which there is the change of slope in the profile, as seen in Fig. ~\ref{figJump}. Instead, we use the ``Equal Deviation'' (ED) method, which we have devised for this purpose. It consists in locating the position of the break at the point in which the inner and outer fit lines, are at a maximum and equal distance (in $\mu$) from the intensity profile. In most cases, there are usually two points of the profile which are at a same distance from the fits. One is the point of intersection of the inner and outer fit lines (a meaningless position in these cases), and the other is the point of interest, the point at which the change in slope appears to happen. By selecting the point at which the distances are equal and of maximum absolute value, the required point is usually naturally selected. Only in some cases, with more irregular profiles, the region in which we search for the points of equal deviation must be limited to find a meaningful solution. If we apply this method to a profile in which there is normal break, i.e. without a ``Jump'', the same solution is obtained by this method as with the standard intersection technique. It is important to note that this method is used only to measure the radius of the break (understood as the change in the slope of the profiles) with greater accuracy. We think that, at this point, and for the stated purposes, it is not necessary to speculate about the causes for these ``Jump Truncations''. Our standpoint is to consider them as irregularities in the profiles which make the task of estimating the position of the ``break'' more involved than in most other cases. In analysing the profiles we adopt the following assumption. A ``break'' in a stellar disk is a significant discontinuity in the slope of the exponential profile. This significance comes from both the change in slope and the permanence of the change at increasing radii. Bearing this in mind, the position of the break should be placed at the point where the change in slope takes place, and not where the lines which best fit the two subsections of the profile intersect, if these two positions do not coincide, as is the case for the ``Jump Truncations''.\\

Taking into account that stellar disks have their irregularities and asymmetries, it might be argued that perhaps other profile analysis techniques would be better fitted to our goals. For example, using averages of profiles in different sectors of the galaxies, instead of the complete azimuthal profiles, might be helpful to indicate problems in individual extracted profiles. Three are the main reasons which make us prefer the adopted methodology. First, our profiles are not azimuthally averaged, but we extract median values of intensity, so minimizing the relevance of asymmetries on the resulting intensity profiles. Second, averaging profiles in sectors would cause a severe decrease in the limiting surface brightness which could be analysed. For many cases, this could imply that we would not be able to measure the truncations. And third, selecting the sectors could add a bias which could compromise the statistical significance of the results.

The complete list of objects (505), with detailed results of the profile classification and characterization, is given in Table \ref{tblResults}. Also listed are the objects which were rejected for analysis. These have no information on classification, and the reason for rejection is given in the last column. We will discuss the results in the following sections.

\section{COMPLETENESS \& STRUCTURAL PARAMETERS RELIABILITY}\label{sec4}

\subsection{Completeness}\label{sec4.1}

\begin{figure}
\epsscale{1}
\plottwo{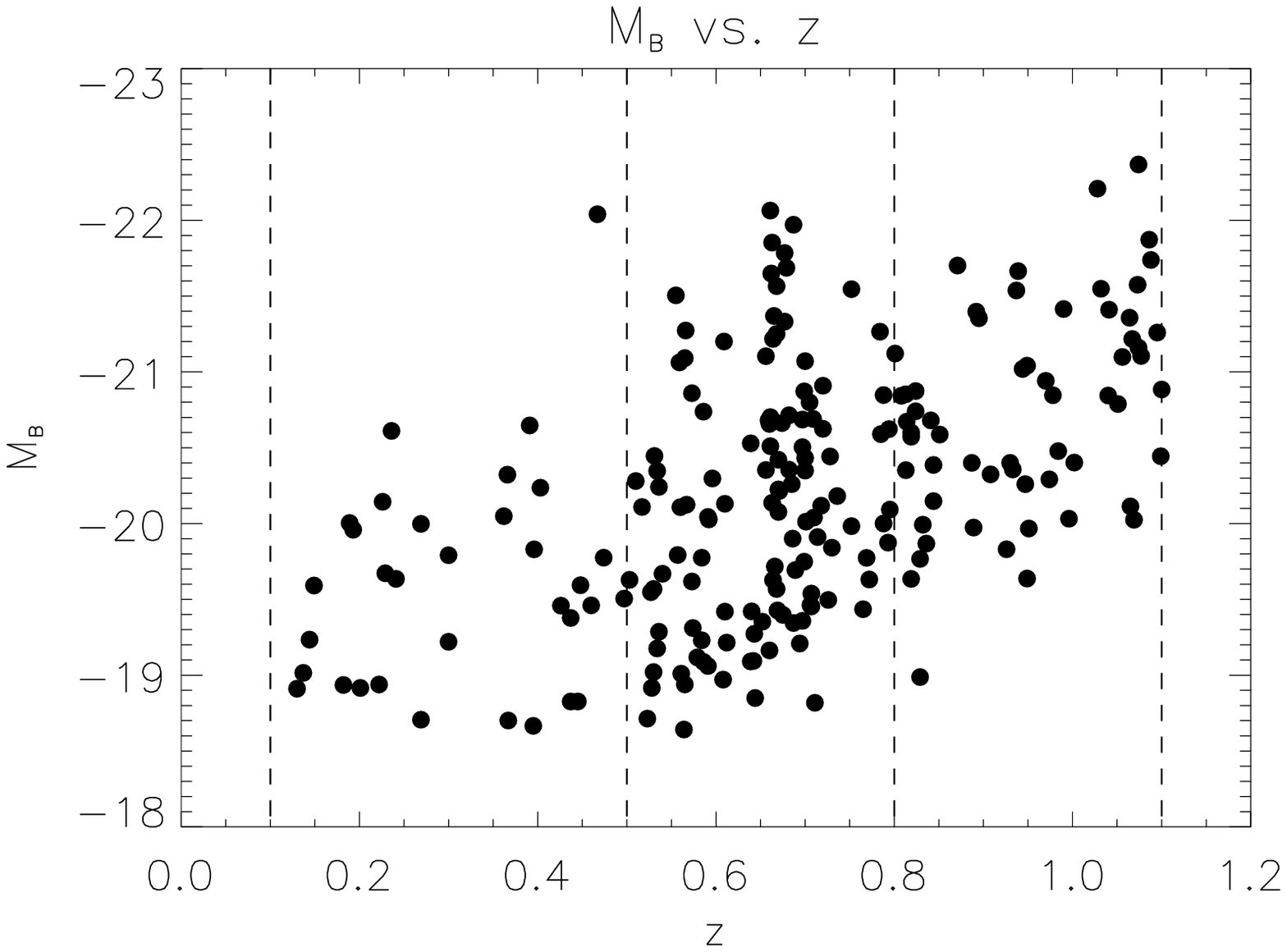}{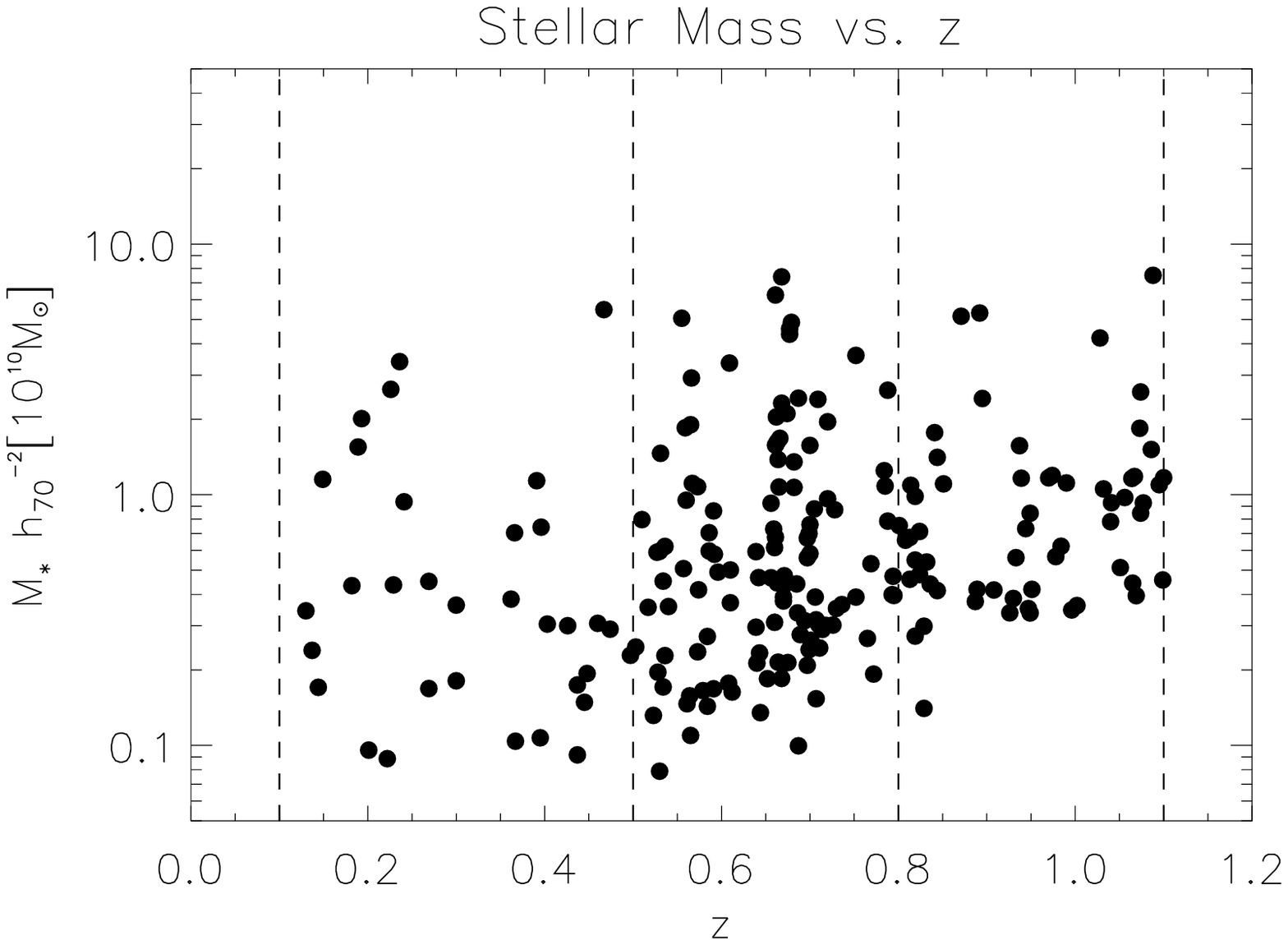}
\caption{Left: Absolute $B$-band magnitude of the galaxies in our sample against their estimated redshift. Right: stellar masses of the same objects against redshift. The vertical lines mark the limits of the 3 redshift bins in which we have divided our sample for study : 0.1$<$z$\leq$0.5; 0.5$<$z$\leq$0.8; 0.8$<$z$\leq$1.1. The completeness levels in absolute magnitude and stellar mass at z$\sim$0.7 are \mb $\sim$ -19.5 mag and \mstar $\sim$ 3$\cdot$ $10^{9}$ \msun respectively.\label{figComplete}}
\end{figure}

The selection criterion employed for the absolute B magnitude (\mb $<$ -18.5) is the same as that used in \citet{TP05}. They adopted this ``cut'' as a compromise between maximizing the number of objects in their final sample and assuring as much homogeneity (equally luminous objects) as possible through the range of redshifts they explored, which is the same as ours. In Fig. \ref{figComplete} (left) we see that at redshift $z$$\sim$0.7, which is the mean/median redshift of our sample, the completeness level is at \mb $\sim$ -19.5 mag, a magnitude below the selected level. This makes our sample incomplete at fainter magnitudes beyond $z$$\sim$0.5. Nevertheless, we decided to maintain our absolute magnitude selection criterion to satisfy the following objectives: a) to allow a detailed comparison with TP05, and b) because otherwise our sample at low redshift would be significantly reduced. It is also interesting to note from the same figure that our sample lacks some high luminosity objects at lower redshifts because the surveyed volume is significantly smaller in that range. In the right panel of the same figure we see the completeness level in stellar mass is \mstar $\sim$ 3$\cdot$ $10^{9}$ \msun at z$\sim$0.7. The incompleteness with stellar mass is less severe than with luminosity. This is because the \mstar/L relation evolves with redshift in this range, in the sense that for a same \mstar galaxies at $z$$\sim$1 are brighter in the $B$-band than at $z$$\sim$0 \citep[eg. ][]{Rudnick03}.

\subsection{Comparison with Radial Profile analysis in TP05}\label{sec4.2}

\begin{figure}
\epsscale{0.9}
\plotone{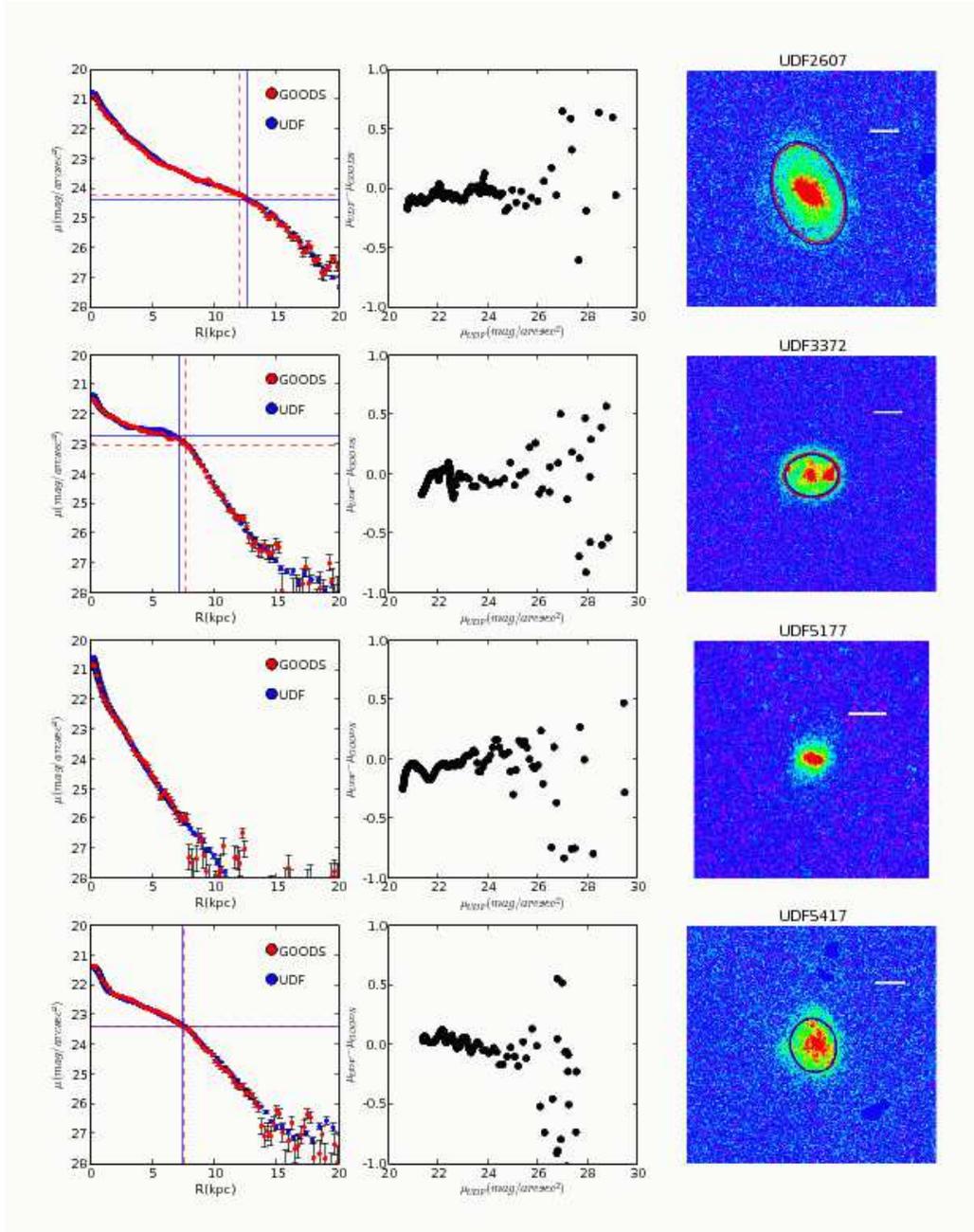}
\caption{Comparison of the surface brightness profiles for 4 galaxies in common to the TP05 sample and ours. The vertical and horizontal lines mark the position of the Break, when there is such, according to each work (in the same color as in the profile). The differences between the two profiles as a function of the brightness meassured in TP05 are shown in the middle panels. We see how these differences grow sharply larger when we reach the $\sim$26 \magarcsq isophote. On the right, images of each object are shown. Ellipses with the Break Radius as semi major axis are also drawn (horizontal white lines are 1'' long). The ellipticities, position angles and centers of these ellipses are our own.\label{figU2G}}
\end{figure}

To check the accuracy of our structural parameter determination and galaxy type classification, we made comparisons with deeper observations and with simulations. Trujillo \& Pohlen 2005 (TP05) studied radial profiles for a sample of 36 late-type galaxies, using imaging data from the \emph{Hubble Ultra Deep Field} \citep[UDF;][]{Beckwith06}, which is also contained in GOODS-South. Those objects are included in our sample, and 35 of them were analysed. Only one object (UDF3203) was discarded in this work, as it has $B$ = -18.46 $>$ -18.5 mag, which is one of our selection criteria. Of the 35 matched objects, 3 more objects were discarded in our analysis: one because it was near the edge of a tile (UDF7556), and the other two as being probable merger candidates (UDF8049 and UDF8275). In Fig.~\ref{figU2G} we show some examples of radial profiles using UDF data and GOODS-S data. We see that the profiles match well ($\vert \mu_{UDF}-\mu_{GOODS} \vert \lesssim 0.2$ \magarcsq) out to a level of $\sim$ 26 \magarcsq, and then the differences grow erratic, as the UDF data are ``deeper'' . We take this value as a minimum surface brightness level of reliability for our profiles.

Of the 32 objects which were in the TP05 sample that we studied, in 22 cases (69\%), we assign the same classification for their profile (as Type I, II or III). The remaining 10 cases (31\% of the matched sample) show varying bases for disagreement. In 4 cases (13\%, UDF3268 II$_{TP05}$-I$_{ATB08}$, UDF3822 III$_{TP05}$-I$_{ATB08}$, UDF9455 II$_{TP05}$-I$_{ATB08}$, UDF6853 III$_{TP05}$-I$_{ATB08}$) the feature which could be taken as a ``break'' can also be identified in our profiles, but happens in TP05 at a lower surface brightness part of our profiles which for our purpose is unreliable (i.e. \mubreak $\gtrsim$ 26 \magarcsq). In 2 cases (6\%, UDF6862 III$_{TP05}$-II$_{ATB08}$, UDF 8257 III$_{TP05}$-II$_{ATB08}$), we have suspicions of neighbor objects being the cause for these galaxies being classified as Type III in TP05. Three galaxies (9\%), present morphologies which are rather irregular or have profiles of difficult interpretation, because of bars (UDF2525 -II+III?- II$_{TP05}$-III$_{ATB08}$, UDF8040 -``Jump Truncation''- I$_{TP05}$-II$_{ATB08}$, UDF7559 -barred- III$_{TP05}$-II$_{ATB08}$). And only in 1 case (3\%, UDF4491 II$_{TP05}$ - I$_{ATB08}$) we can see no feature in our profiles that may justify the difference in classification. More importantly for this paper, in the case of Type II galaxies there were only 2 cases out of 19 ($\sim$10\%) that were misclassified due to an insufficient depth of the GOODS images.

\begin{figure}
\plottwo{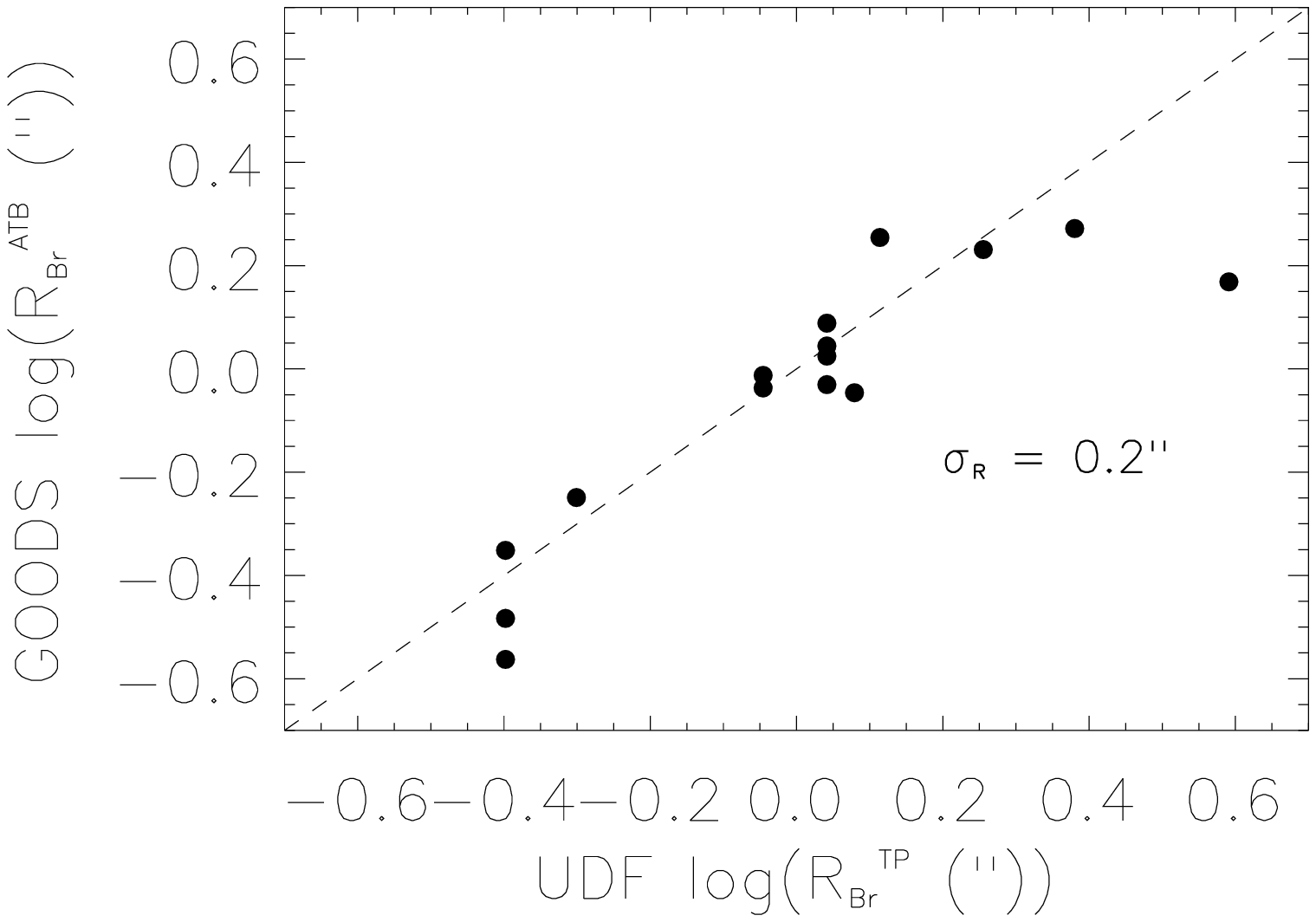}{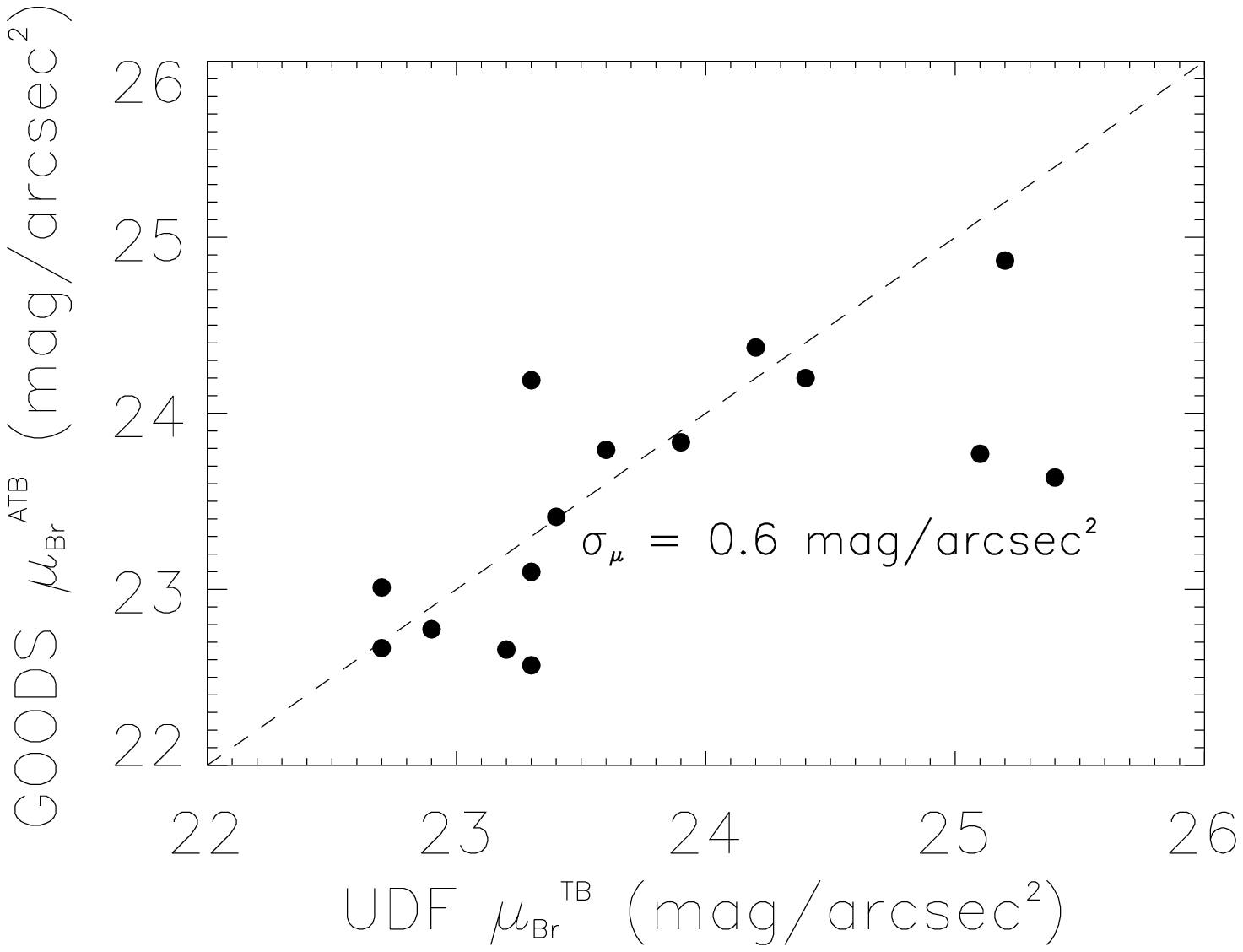}
\caption{Comparison of estimates of the Break Radius (\rbreak; left) and Surface Brightness at the Break (\mubreak; right) for a sample of 15 galaxies (Type II), common to this work (ATB08) and to TP05. The discontinuous lines represent a 1:1 relation.\label{figU2G_RMU}}
\end{figure}

In Fig. \ref{figU2G_RMU} we present a comparison of the estimates of the \rbreak and \mubreak for the 15 objects of Type II where our classification agrees with TP05. We see an overall good agreement for both parameters. We define the relative error in \rbreak as 100 * [($R_{TP05}$-$R_{ATB08}$)/ $R_{ATB08}$], and the relative error in intensity as 100 * [1 - $10^{-0.4 (\mu_{TP05}-\mu_{ATB08})}$]. Now, for the given comparisons, the standard deviations are $\sigma_{R}$ = 0.2'' ($\sigma_{R}^{err}$ = 22\%) and $\sigma_{\mu}$ = 0.6 \magarcsq ($\sigma_{I}^{err}$ = 43\%) for \rbreak and \mubreak respectively. For \rbreak, the largest disagreement is with UDF1971. The feature we identified as the break was taken by TP05 as a mere consequence of the bar, while the truncation they detect falls below our reliability level, though it can be seen in our profile. In \mubreak the differences are below 1 \magarcsq in all cases, except for UDF6974 and UDF1971. There is a hint, although based on poor statistics, that breaks at \mubreak (TP05) $>$ 25 \magarcsq are overestimated in brightness in our work relative to TP05 (\mubreak(ATB08) $<$ \mubreak(TP05)).

\subsection{Profile Analysis of Artificial Galaxies.}\label{sec4.3}

In addition to the previous comparison, we performed simulations in order to assess the accuracy of our results, with regard to the classification of the profiles and their characterization. We have created 300 synthetic galaxies using the GALFIT program \citep[][see the Appendix for a detailed description of how the objects were created and analysed]{Peng02}. The mock objects were placed in a synthetic image, to emulate the signal-to-noise ratio, angular scale and PSF properties of the \zz band images of GOODS-S. The synthesized galaxies are exponential disks of Type I, II or III (100 of each kind). Their total apparent magnitudes, inner scale length \hone over every Type, and outer scale length \htwo, \rbreak and \mubreak for Types II and III, are taken to cover the whole ranges over which those parameters vary for our real objects. The selected ranges of parameters are in Table \ref{tblSim}.

The simulated objects were detected with SExtractor and analysed in an analogous way to that used in studying the real objects. Each model was visually classified as Type I, II or III. When the object had been assigned to Type II or III, the position of the break in the \mur plane was also estimated. The process was done ``blindly'', the classifier did not know which were the real classification and profile parameters of the object a priori.

\begin{figure}
\epsscale{1}
\plottwo{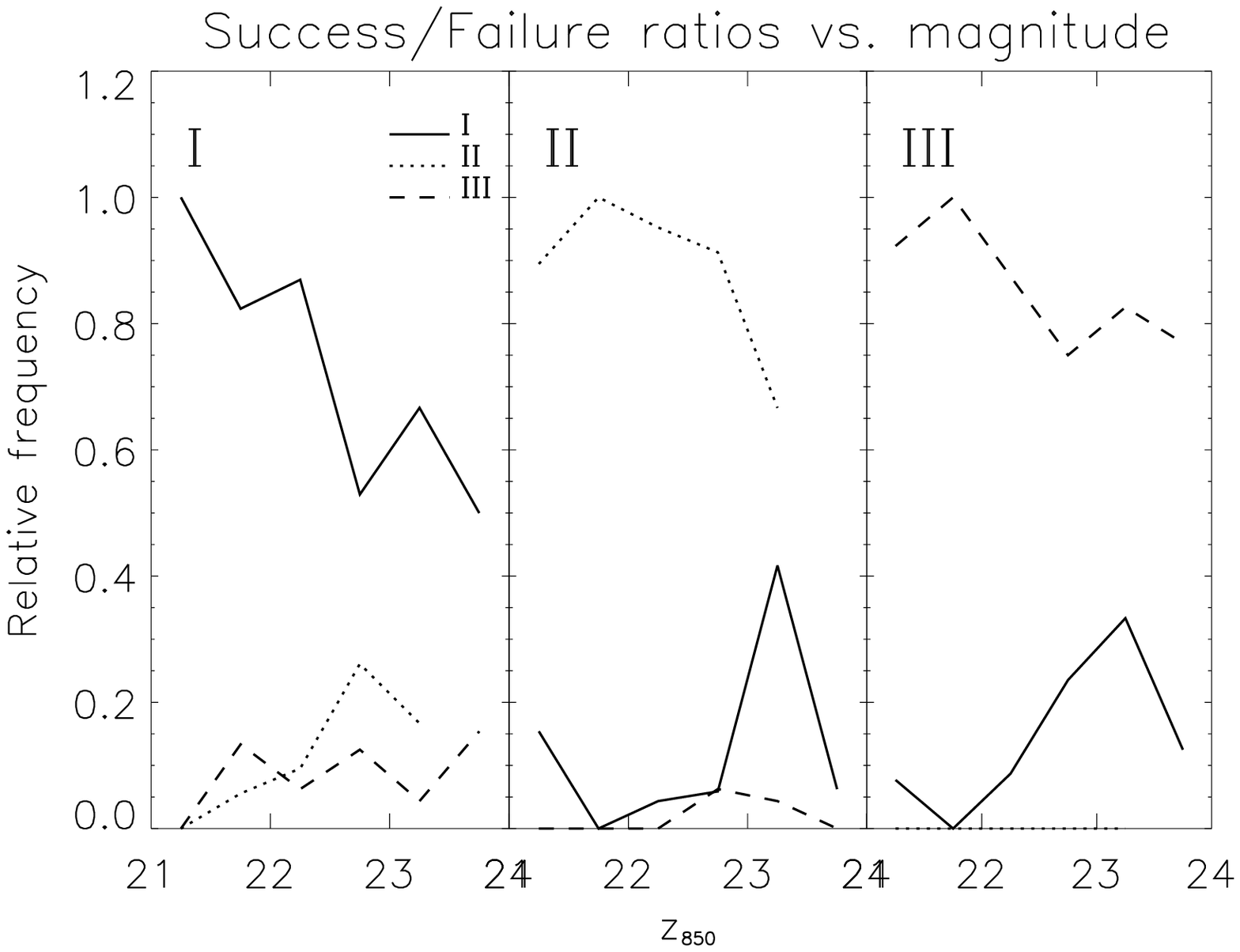}{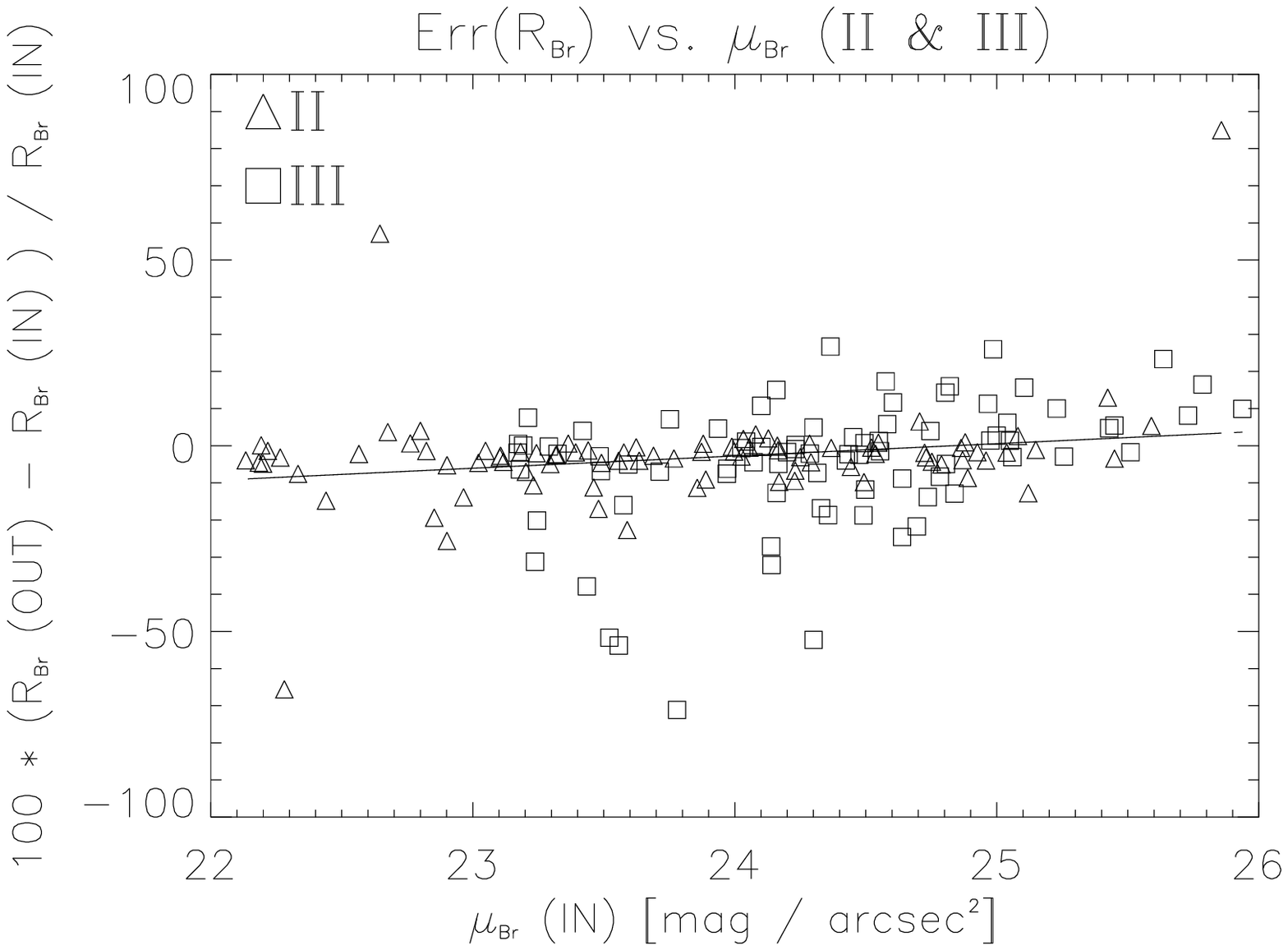}
 \caption{Results for simulated galaxies: left) classification success ratio and fraction of objects which are misclassifed as being of the other 2 types as a function of the \zz magnitude of the simulated galaxy; right) relative error in the visual estimate of \rbreak, as a function of the input surface brightness at the ``break'' \mubreak$_{in}$, for synthetic Types II and III (coded with different symbols). The uncertainty of the break position is $\sigma_{err}^{II}$ = 5\% for Type II objects, and $\sigma_{err}^{III}$ = 13\% for Type III. For both Types together, it is $\sigma_{err}$ = 8\%.}\label{figSim_mag}
\end{figure}

In Fig. \ref{figSim_mag} (left) we show the effect of the apparent magnitude of the object on its classification. In the panels we have plotted the fraction of objects of each Type (I, II or III) which are correctly classified, against the total apparent magnitude. In each panel the lower 2 curves represent the fraction of objects which are misclassified as being of the complementary Types. Only points which represent a population of at least 3 objects are shown. In the first panel we see how the success ratio in classifying Type I objects goes from 100\% to 50\% at the fainter end. The wrong classifications are divided between Types II and III with roughly equal shares. In the second panel we see how Type II objects are classified with a higher ratio of success at all surveyed magnitudes ($\gtrsim$70\%), but fainter objects are principally misclassified as being of Type I. Finally, as shown in the third panel, Type III ($\gtrsim$80\%) objects are never misclassified as Type II, but Type I. It is interesting to note that the reliability of our classification, both in models and real galaxies (Sect. \ref{sec4.2}), are at a similar level: 70-80\%. Furthermore, note that more than 80\% of our real galaxies have \zz$<$23 mag, for which the probability of a potential misclassification is lower.

As an illustrative exercise, we have also tested the reliability of the visual appreciation of the position of the ``break'' in the simulations. In Fig. \ref{figSim_mag} right we represent the relative error in the estimate of \rbreak, against the input surface brightness at the ``break'' \mubreak$_{in}$. The relative error in \rbreak is defined as 100 $\cdot$ (\rbreak$_{out}$ - \rbreak$_{in}$) / \rbreak$_{in}$, where \rbreak$_{in}$ is the actual position of the break for the model and \rbreak$_{out}$ is the visually estimated position. In the Figure we plot the results for Types II and III. The points have been fitted by a line, and this gives a variation in the relative error from -9.4\% to 3.9\% between 22 and 26 \magarcsq in \mubreak$_{in}$. The standard deviation in the relative error is $\sigma_{err}$ = 8\%. We see how the scatter for Type II objects is less ($\sigma_{err}^{II}$=5\%) than for Type III objects ($\sigma_{err}^{III}$=13\%). These errors are both below the estimate of the error in \rbreak (22\%) derived from the comparison of results in this work and TP05 for objects common to GOODS and UDF observations, as reported in subsection 4.2. It is that error, derived from the analysis of real galaxies, which must be taken a realistic estimate of the error in the estimate of \rbreak.

\begin{figure}
\epsscale{1}
\plotone{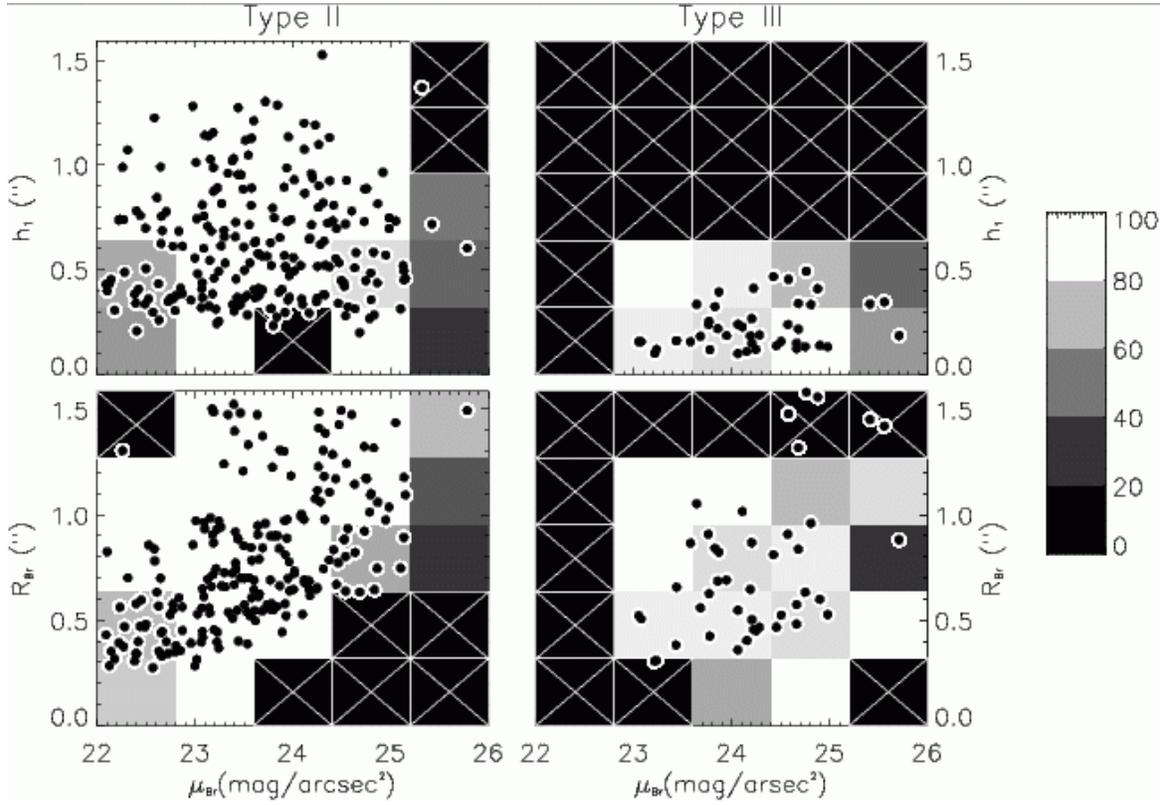}
\caption{Probability map of success in the classification of simulated galaxies, as a function of simulated \hone and \mubreak (up), and \rbreak and \mubreak (down). We present results for input models of Type II (left) and III (right). The points represent the position of real objects of our sample in the given planes. The 'crosses' mark the bins for which there were no input models available in the simulation. Note how the vast majority of Type II real galaxies are in a portion of the plane where the success classification is estimated to be very high ($>$80\%).\label{figHisto2Dsim}}
\end{figure}

Finally, in Fig. \ref{figHisto2Dsim} we present probability maps of the success in classification of simulated objects of Type II (left) and III (right), as a function of the simulated \hone and \mubreak (up) and \rbreak and \mubreak (down) of the models. The success ratio has been coded in gray scale, and the portions of the planes (\hone - \mubreak and \rbreak - \mubreak) for which no input model within the corresponding ranges of parameters was produced are marked with a cross. The points mark the position in the given planes of the real galaxies under study. Most of our real galaxies are in a position of the plane where the classification success is $>$80\%.

To summarize the results of this section, from a comparison with deeper observations (i.e. with the UDF) and with artificial galaxies in simulations we conclude that for Type II galaxies (the main goal of this paper) the accuracy on type profile identification is higher than 80\%, for surveys as deep as GOODS-South. Nearly 20\% of Type II galaxies that are potentially missed in our work are not found because their break positions lie at fainter magnitudes than 25-26 \magarcsq.

\section{Results}\label{sec5}

\subsection{Classification of Galaxies in our Sample.}

We have divided our sample in 3 redshift bins or ranges; ``low'': 0.1$<z\leq$0.5, ``mid'': 0.5$<z\leq$0.8 and ``high'': 0.8$<z\leq$1.1. Of the 435 objects which were suitable for analysis, 146 (34\%) are classified as Type I, 242 (56\%) as Type II and 47 (11\%) as Type III. All considered objects are visually confirmed as non Irregular/Merger, as we stated before. There are 6 objects, of Types I and III, which are suspected of being of early-type (probably S0, though it is difficult to be sure without more information), but have entered our statistics. These objects are marked as ``Early'' in Table \ref{tblResults}, and the reason to include them is that they seem to have a genuine disk. Nonetheless, and to avoid any controversy on this issue, our results on disk sizes are based on a subsample of the Type II objects, of which none is classified as ``Early''.

Our work is focused on the population of late-type, non interacting galaxies, in which the profile is ``externally'' truncated. This means we concentrate on Type II objects, and specifically on those in which the ``downbending'' break takes place in the outer parts of the visible disk, i.e. beyond the visible bars or spiral arms. A selection criterion for the ratio \rbreak / \hone, where $h_{1}$ is the scale length of the inner part of the disk, is not effective in discriminating between the two kinds of ``breaks'': the ``inner'', which take place ``inside'' the stellar disk, and the proper ``truncations'', those which mark the edge of the stellar disk, or a decline in the density of the stellar population, and so the trained judgement of the classifier is needed to distinguish between them. We define, then, our ``truncated'' sample (T-sample hereafter) as follows: late-type objects, though not Irregular, nor involved in merger events, with a Type II-truncated profile. This sample contains 238 objects. That means only 4 out of 242 objects which form the whole Type II sample have breaks of the ``inner'' kind. Additionally, 48 of the objects in the T-sample have asymmetries in their disks (20\%). Finally, there are a few objects (15, 6.3\%) in which the profile is of the mixed kind II+III. In these there is a clear Type II break, followed by a second break, of Type III; even so we have included them in the T-sample.

The distribution of the objects amongst the 3 redshift ranges, is given in Table \ref{tblClass}. We find that the frequency of objects classifed as of Type I decreases from 39\% to 25\% between z$\sim$1 and z$\sim$0.3. For Type II objects the opposite trend is found, increasing their proportion by almost 9\% in the same period. These results are compatible with no changes in the distributions within the error bars. Finally, Type III objects show no change in their relative population within the error bars between z$\sim$1 and z$\sim$0.3. It is tempting to explore whether there is a real evolution in the populations of objects of different profile types. This would mean that more disk galaxies presented no truncations in the past than nowadays, relative to the fraction of those that present truncation. But first we need to look again Fig. \ref{figSim_mag} (left), in which we show results from the simulations. As we stated above, when Type I objects are fainter, they are increasingly mistaken for Types II or III in roughly equal proportions. On the other hand, genuine Type II objects are increasingly classified as Type I when the magnitude decreases. Consequently, it is quite possible that the observed variation in the frequencies is caused by misidentifications of Type II galaxies as Type I at high $z$.

We also compare the given ratios of profile types with results published in PT06 for local disk galaxies in the SDSS (their sample being composed of 85 objects, with Hubble Types $T$ in the range 3$<$T$<$8.5). In the Table 4 of that paper they report the following shares: 11$\pm$3\% of objects are classified as Type I, 66$\pm$5\% objects have a Type II profile and 33$\pm$5\% have a Type III profile. The shares do not add up to 100, and that is because they count objects with a mixed classification (i.e. II+III or III+II) for the statistics of both Types. In our lowest redshift bin (z$\sim$0.3) the objects are classified as 25$\pm$6\% of Type I, 59$\pm$9\% of Type II and 15$\pm$5\% of Type III. We see the fraction of Type II objects is in good agreement with their results, while there is significant disagreement with regard to Types I and III. We assume that the fact that Type III is under represented in our work, in comparison with theirs, is probably because Type III objects have their breaks at lower brightness levels than Type II's, and so are harder to identify at intermediate redshifts.

\subsection{Results for the ``truncated'' sample.}

As stated in Section 2, the classification and characterization of the profiles has been performed in the band which best approximates the rest-frame $B$-band in each redshift bin. As a consequence, the contribution of luminous young stars to the retrieved surface brightness profiles is significant. In fact, in many objects the signature of star forming ``clumps'' spread over the disks is evident in the images. This means that the reported ``truncations'' should be more accurately interpreted as related to, or at least influenced by, the extent of the ``star formation'' disk, a term we will use to refer to that part of the disk where most of the massive star formation is taking place in the galaxies. In contrast, the ``truncations'' found at longer wavelengths in other works (e.g. PT06) could be more unambiguously interpreted as an abrupt decrease in the density of stellar mass. Nonetheless, when comparing profile parameters of objects in the Local Universe (from results in PT06) and at intermediate redshift from this work, we do so based on profile characterizations performed in similar rest-frame bands: $g'$ for local objects and $B$-band for the higher redshift ones.

\subsubsection{ \rbreak - B-Luminosity relation.}

\begin{figure}
\epsscale{1}
\plottwo{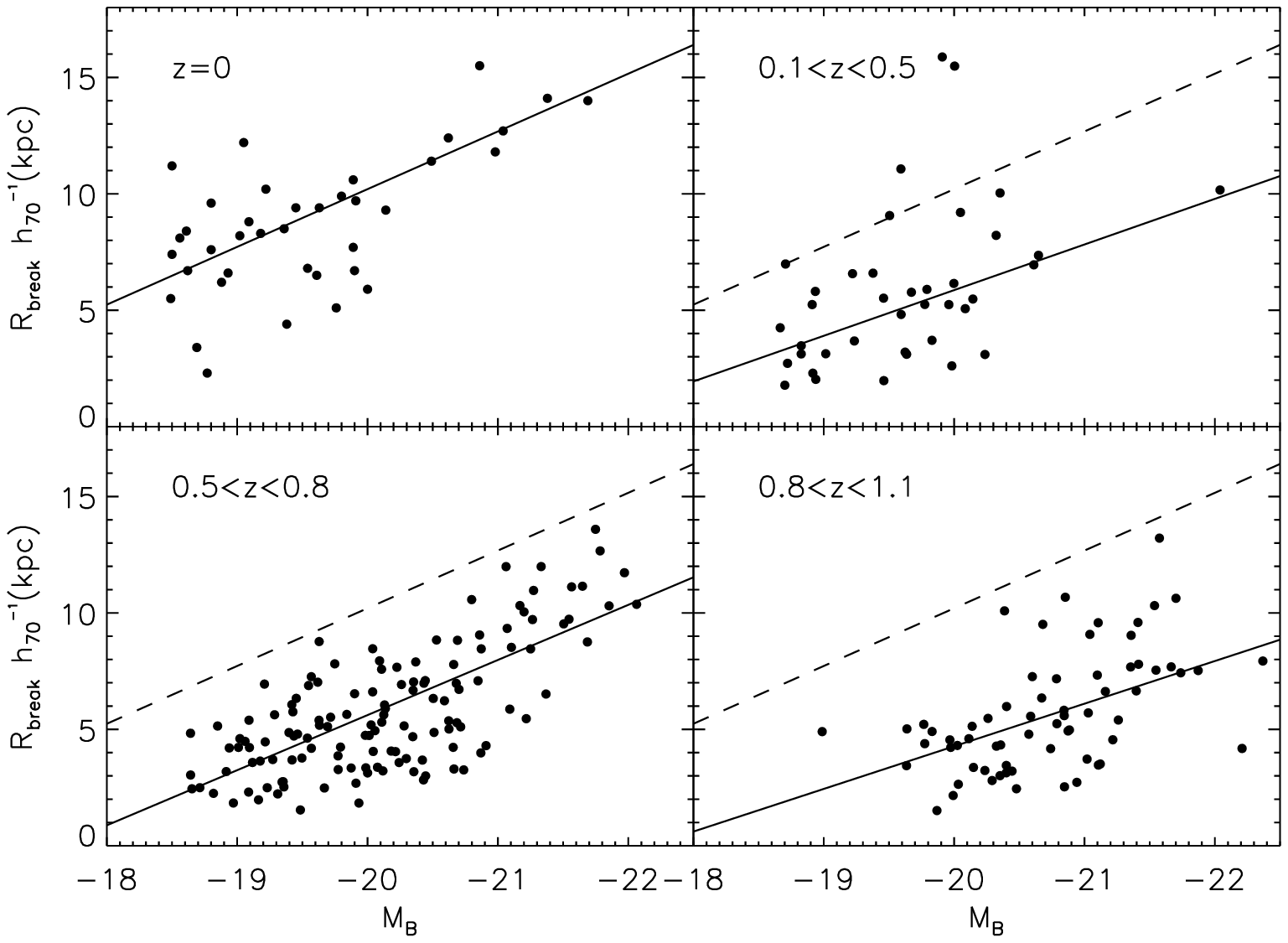}{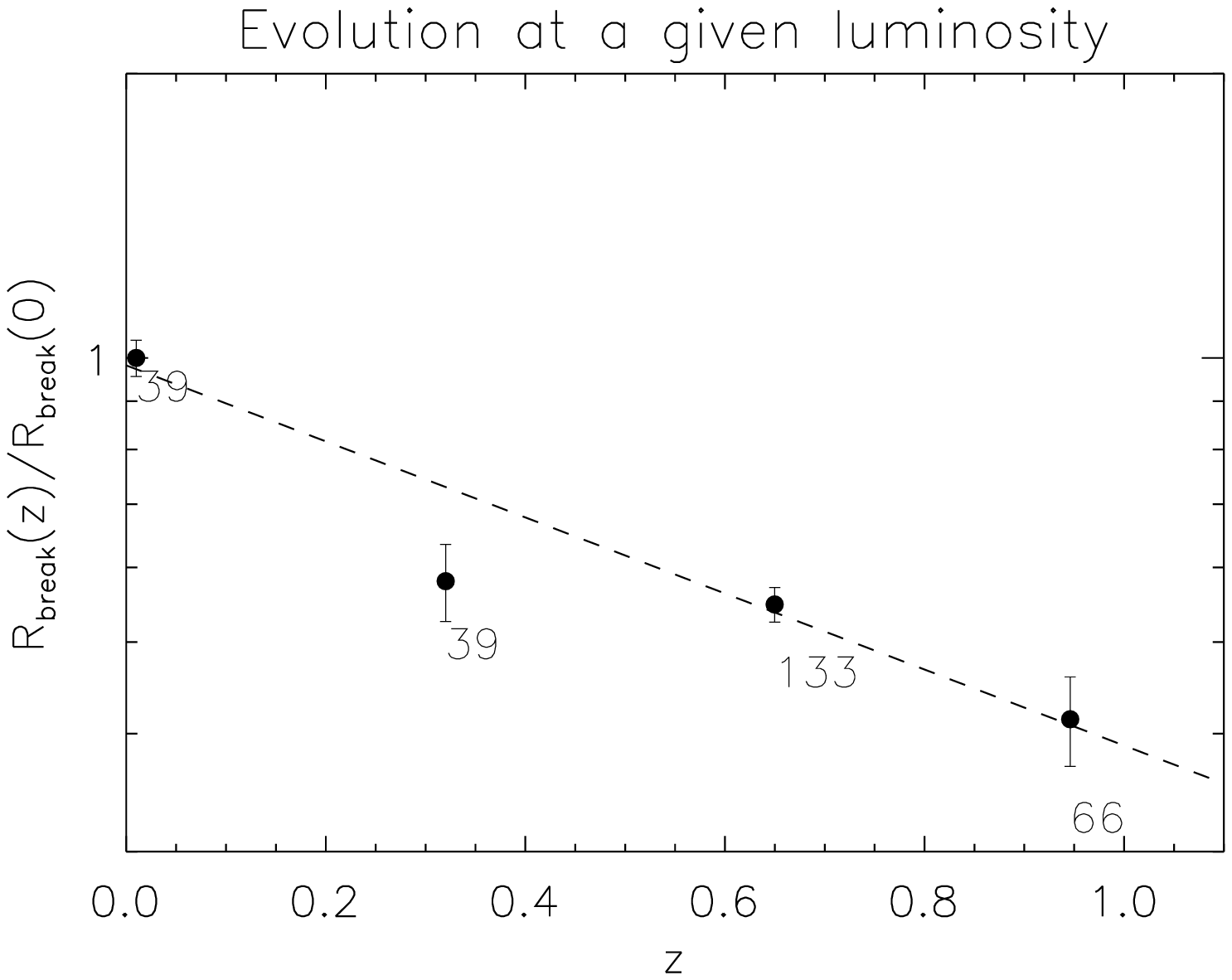}
\caption{Left: Break Radius (\rbreak) for ``truncated'' galaxies as a function of absolute $B$-band magnitude \mb, for 4 ranges of redshift. The data for z=0 are from PT06 ($g'$-band results). Right: \rbreak at \mb=-20 mag, given by the best fit lines in the left panels, relative to z=0, for each redshift range. Ordinates are in logarithmic scale. A linear fit to log(\rbreak(z)/\rbreak(0)) - redshift gives a growth in \rbreak by a factor 2.6$\pm$0.3 between z=1 and z=0 for a same \mb. The discontinuos straight line represents the relation for the local reference sample (the z=0 panel). The same applies to Figs. \ref{figTrmass}, \ref{figTrh1mag}, \ref{figTrh1mass}, \ref{figh1mag}, and  \ref{figh1mass}. The numbers acompanying each point in the right panel give the population of objects which they represent.\label{figTrmag}}
\end{figure}

Using the radial position of the truncation as a direct estimator of the size of the ``star formation dominated disk'', we have explored the relation between \rbreak and the $B$-band Luminosity of the galaxies for the T-sample (``truncated'' profiles). In Fig. \ref{figTrmag} left we show \rbreak in kpc against the rest-frame absolute $B$ magnitude \citep[from][]{Barden05}. In the 4 panels we show this relation at the 3 redshift ranges we have explored, and also in the Local Universe, based on results from PT06 on SDSS galaxies at $z$$\sim$0 (we use the values of \rbreak obtained in $g'$-band for these objects). The observed distributions have been fitted to a line with a ``robust'' least absolute deviation fit, in an iterative ``bootstrap method'' to get the most probable values for the slope and the y-intercept ordinate and their errors (the same mathematical procedure is used in Figs. \ref{figTrmass} and \ref{figTrh1mag} through \ref{figh1mass}). At a given luminosity, \rbreak is evolving towards smaller values as redshift increases.

Before exploring the meaning of this last assertion we want to clarify an important point with regard to the distribution of objects with luminosity at different redshifts. In the panels we see how this distribution varies amongst redshift bins. In the ``low'' redshift bin the objects are concentrated around (-20.0, -18.5) mag, while in the ``mid'' range the distribution is quite homogeneous between -18.5 $\leq$ \mb $\leq$ -22 mag. Quite opposite to what happens at lower redshifts, in the ``high'' redshift bin the objects cluster around (-20.0, -21.5) mag. These distributions are the result of a) probing a smaller volume in the lowest redshift bin and b) the sample is incomplete at the lower end of luminosities in the higher redshift bin. To avoid as much as possible the sampling effects just described, we explore the size evolution using as reference point \mb = -20 mag, which is well populated of galaxies at all redshifts.

In Fig. \ref{figTrmag} b) we show the value of the aforementioned best fit lines at \mb = -20 mag, relative to the corresponding value in the $z$=0 sample (logarithmic y-axis), against the mean value of redshift in each bin. The errors shown are from the bootstrap method applied to the fitting of the relation. We see how the ratio decreases with redshift. Given the fact the points are aligned we fit them with a straight line. On this line, the \rbreak of galaxy with \mb = -20 mag has increased by a factor 2.6$\pm$0.3 between $z$=1 and $z$=0. We see how the point at $z$=0.3 fits somewhat worse to the straight line than the others. Nonetheless, the deviation is 2.5$\sigma$.

As we explained above, the ranges of \mb covered by the objects are not the same amongst different redshift bins, because of a lack of completeness inherent to the available data. Although our analysis method has been devised to minimize the negative consequences this fact could have on the produced results, it would be desirable to test, within the possibilities granted by the available data, whether this method is really effective for the stated purpose. In this line, we have also explored the relation between \rbreak and \mb when a same range of \mb is used to select the objects in all redshift bins. This range has been chosen as \mbrange. When applying this restriction, the sample populations reduce to 15, 22, 75 and 41 objects in the z$\sim$0, and the ``low'', ``mid'' and ``high'' redshift bins respectively. In comparison, the whole samples explored in Fig. 7 (and also in Figs. 8 through 14) are 39, 39, 133 and 66 objects in the same redshift bins, i.e., significantly larger. The growth factor in \rbreak between z$\sim$1 and z$\sim$0, deduced from the best fit lines to the relation \rbreak-\mb at \mb=-20 mag is, when restricting the range in \mb, 2.0$\pm$0.8, which is smaller than the value given above, but in agreement within the error bar. This further supports the hypothesis that the reported change in \rbreak at a fixed \mb across redshift is not due to differences in the luminosities range of the sampled objects. 

\subsubsection{ \rbreak - Stellar Mass relation.}

It is not entirely straightforward to interpret the size evolution of galaxies from the \rbreak - \mb relation, studied above, for the following reason. The Mass Luminosity relation, specially for shorter wavelength bands such as the $B$-band, has varied significantly between $z$=0 and $z$$\sim$1 \citep{Brinchmann00,Bell03,Dickinson03,PerezGonzalez08}. Objects of the same mass were brighter in the $B$-band in the past as their stellar populations were younger on average. This has as consequence that even if there were no changes in the \rbreak of objects, their proved increase in $B$-band Luminosity would make the relation vary such that objects of a same \mb would have smaller values of \rbreak in the past. This is the trend reported in the preceding subsection. Fortunately there is a way to overcome the consequences of this luminosity evolution in order to test for a ``real'' change in the sizes of stellar disks with time (or more properly, the size of the ``star formation dominated'' disk, as we are relying on rest-frame $B$-band data). This is to survey the \rbreak - \mstar relation. It is also known that the stellar mass content of galaxies has increased since z$\sim$1, but this change has been more moderate, in relative terms, \citep[$\lesssim$30\%; ]{Rudnick03} than that in luminosity.

\begin{figure}
\epsscale{1}
\plottwo{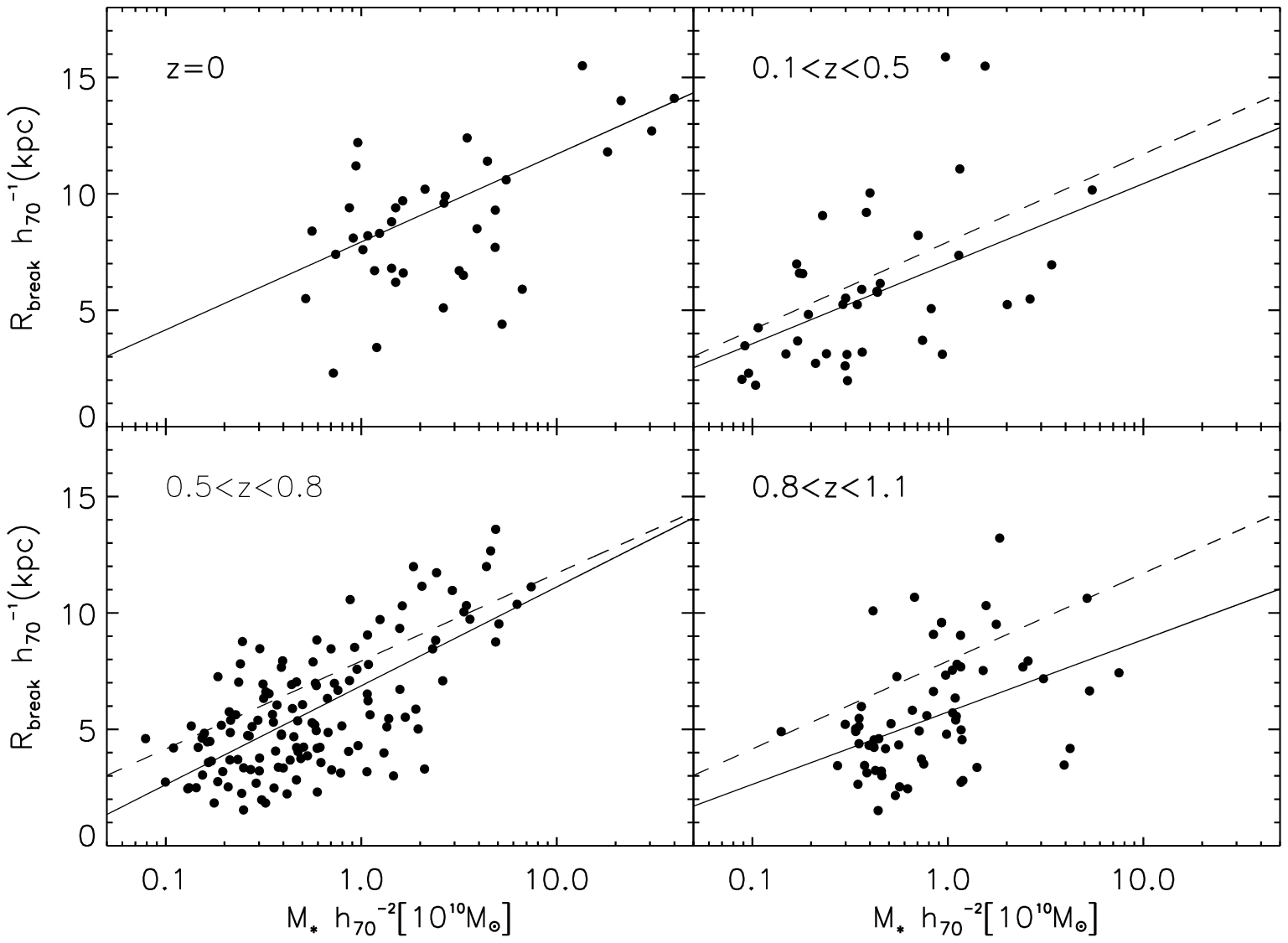}{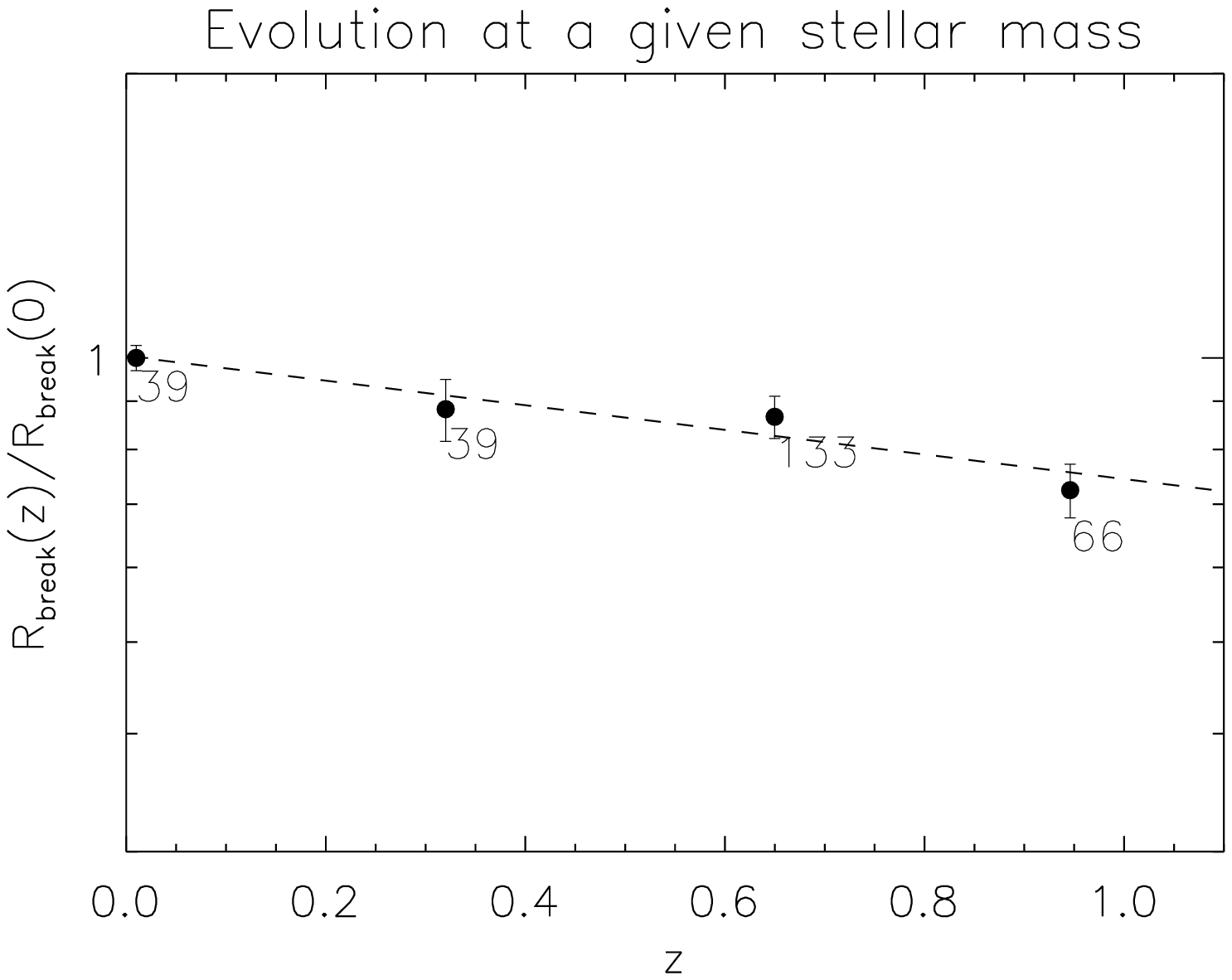}
\caption{Left: Break Radius (\rbreak) of ``truncated'' galaxies as a function of stellar mass \mstar, for 4 ranges of redshift. Local data are from PT06 ($g'$-band results). Right: \rbreak at \mstar=$10^{10}$\msun, given by the best fit lines in the left panel, relative to z=0, for each redshift range. Ordinates are in logarithmic scale. A linear fit to log(\rbreak(z)/\rbreak(0)) gives a growth in \rbreak by a factor 1.3$\pm$0.1 between z=1 and z=0, at a given \mstar. The numbers acompanying each point in the right panel give the population of objects which they represent.\label{figTrmass}}
\end{figure}

Following the preceding discussion, we have tested the relation between size (\rbreak) and stellar mass. This is the strategy adopted in \citet{Trujillo04} and \citet{Trujillo06} to test for evolution of the effective radius of massive galaxies (\mstar $>$ $10^{10}$ \msun) at high redshifts. In Fig. \ref{figTrmass} we show, in analogous fashion to Fig. \ref{figTrmag} left, the values of \rbreak for objects of the T-sample against the stellar mass \mstar, as reported in \citet{Barden05}. The $z$=0 data come again from \citet{TP05} \citep[their stellar masses were computed following the prescription of][]{Bell03}. In this case, the best fit line (obtained by the same method as in the \rbreak - \mb relation) also does show a trend to lower values with redshift. While in this case the differences in the distribution of stellar masses are not as large as in the case of the luminosities, we see how the slopes are slightly less ``stable'' than in that case, based on the assumption that they should follow less pronounced changes between redshift ranges. This may be because the distribution of masses is not as broad relative to the distribution of \rbreak as the distribution of \mb's. Moreover, the stellar mass is a quantity derived from the luminosities, and this leads to an increase in the dispersion. The factors that go into producing the dispersion in the two relations (\rbreak - \mb and \rbreak - \mstar), are the following: a) inaccurate redshifts ($\delta$$z$/(1+$z$)$\sim$0.02), b) inaccurate estimates of the \rbreak ($\sigma^{err}_{R}$ $\lesssim$ 25\%), and c) intrinsic dispersion in the relation. Though the two first factors (especially the second) may play a role in explaining part of the dispersion, we do not rule out the third possibility; i.e. it is not only possible, but almost to be expected, that objects of the same mass may have followed different evolutionary paths (depending on initial conditions and environment, for example), which have shaped their structure in a different way, and so produce ``intrinsic'' dispersions in the \rbreak - \mb and \rbreak - \mstar relations.

In Fig. \ref{figTrmass} right we present the fitted value of \rbreak at $10^{10}$ \msun relative to the $z$=0 value, against the mean value of redshift in each bin. The errors are again derived from the bootstrap method applied when fitting the \rbreak-\mstar relation. We see a less strong evolution with redshift than for the \rbreak-\mb relation. It is also interesting to see how in this case the points are much better aligned than for that relation.

From the linear fit to the points in Fig. \ref{figTrmass} right we deduce an increase by a factor of 1.3$\pm$0.1 in \rbreak for a given stellar mass between $z$=1 and $z$=0, i.e. in the last $\sim$8 Gyr using the standard parameters in our cosmological model. For comparison, in the same range of redshifts, and using the assumption of a maximum evolution in luminosity for their galaxies, \citet{TP05} reported a more moderate growth in \rbreak by a factor of at least 1.25, which is in good agreement with our result. 

Another point worth noting is that individual galaxies also evolve in stellar mass. For this reason it is interesting to provide a rough number of how much \rbreak could grow for a given ``individual'' galaxy. In the range of ages probed here it is claimed in the literature that galaxies have increased their stellar masses by $\sim$30\% \citep{Rudnick03}. This means, that a Type II object with \mstar = $10^{10}$ \msun at z=1, would have its \rbreak larger by $\lesssim$50\% today, i.e. by somewhat more than if its mass were unchanged. 

We have also explored the relation between \rbreak and \mstar when a same range of \mstar is imposed to select the objects in all redshift bins, in analogous way to what we did in subsection 5.2.1 for the \rbreak-\mb relation, and for the same reasons. In this case, the chosen range of stellar masses is \mstarrange. In this case the populations reduced to 30, 11, 59 and 40 galaxies within the z$\sim$0, and ``low'', ``mid'' and ``high'' redshift bins, respectively. As a result, and for these stellar mass-restricted samples, the growth factor in \rbreak between z$\sim$1 and z$\sim$0 is 1.4$\pm$0.2, a minor difference with the value obtained for the unrestricted samples. This difference between ``restricted'' and ``unrestricted'' values is lesser than that found for the \rbreak-\mb relation because the distributions of \mstar are more similar between the redsfhift bins than the distributions in \mb. Finally, this test strengthens the significance of the result found for the growth in the \rbreak-\mstar relation, as it persists when only objects within a same stellar mass range are taken into account through redshift.

\subsubsection{Surface Brightness at the Break Evolution.}

\begin{figure}
\epsscale{1}
\plottwo{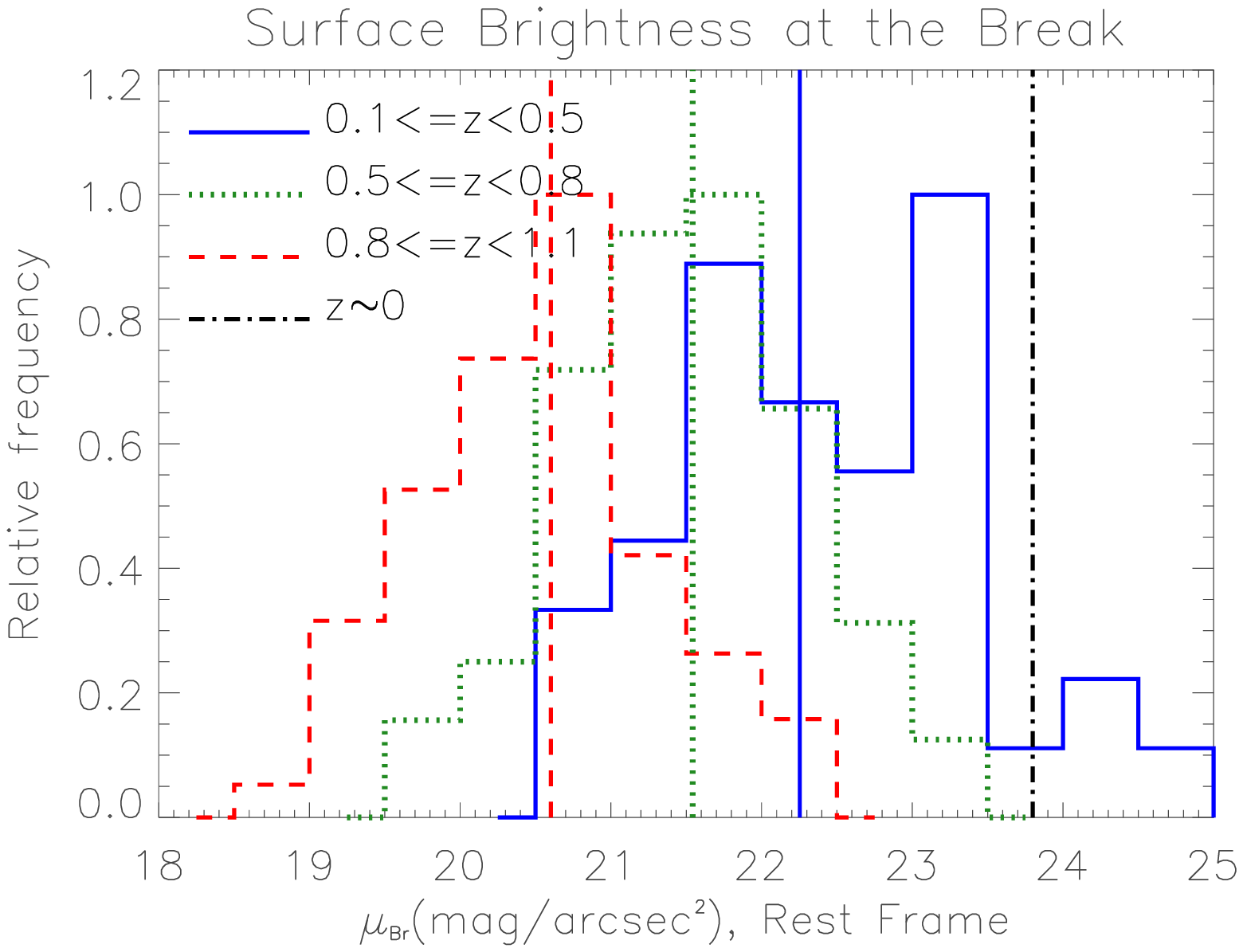}{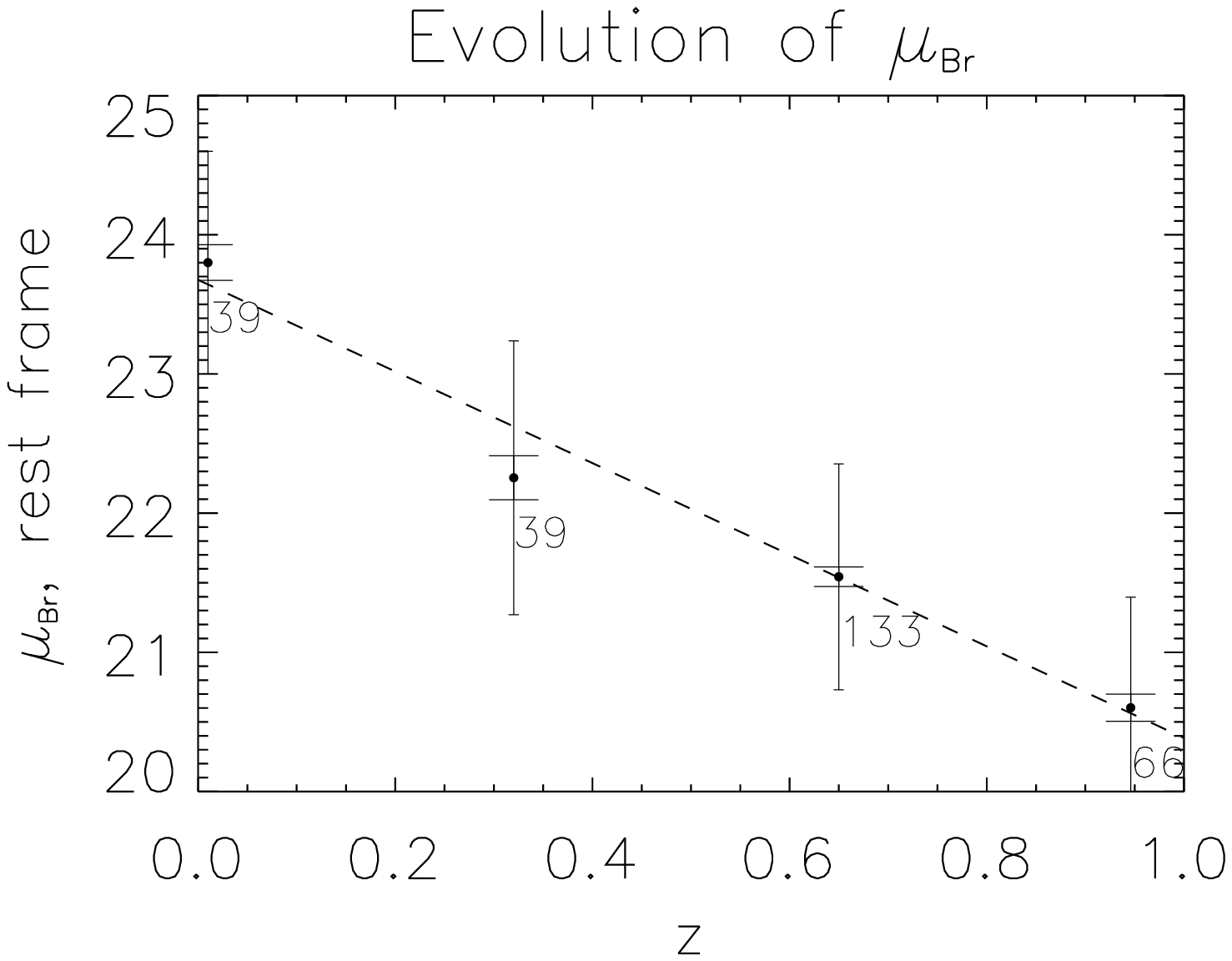}
\caption{Evolution of the Surface Brightness at the Break (\mubreak), in ``truncated'' galaxies, with redshift. Left: the \mubreak's are rest-frame, i.e., they have been corrected for cosmological dimming (I(z) $\propto$ (1+z)$^{-4}$). From z$\sim$1 to z$\sim$0 the median in the distributions of \mubreak has decreased by 3.3$\pm$0.2 \magarcsq (a factor 20.9$\pm$4.2 in intensity). The median values are represented as vertical lines. For comparison we also show the median of the \mubreak values given in \citet{PT06} for local galaxies, in g' band: \mubreak$^{g', PT06}$ = 23.8$\pm$0.8\magarcsq. Right: Median value of \mubreak for the distributions shown in the left panel, against their mean redshift. The larger error bars represent the standard deviation of the distributions, while the shorter ones give the error in the median values. The reported increase of 3.2$\pm$0.2 \magarcsq in \mubreak with redshift is computed from the linear fit to the points. The numbers acompanying each point in the right panel give the population of objects which they represent.}\label{figMuz}
\end{figure}

We have also explored the distribution of \mubreak in our sample of truncated profiles. In Fig. \ref{figMuz} (left) we show a histogram of the \mubreak distributions of the T-sample in the 3 redshift bins under study. The \mubreak have been corrected for the cosmological dimming effect (I $\propto$ $(1+z)^{-4}$). This means we are showing the distribution of \mubreak as they would be measured from a rest-frame observational standpoint for every object. Also in the same panel we represent the median of the distribution of \mubreak in the $g'$ band reported in PT06 for $z$$\sim$0 galaxies. We use this band as it is the closest to our rest-frame $B$-band. We see a clear evolution in the median values of the distributions, in the sense that the ``break'' in the profiles happens at a brightness level which is 3.3$\pm$0.2 \magarcsq brighter at $z$$\sim$1 than at $z$$\sim$0. This is a strong evolution, by a factor of 20.9$\pm$4.2 in intensity. As we are measuring the surface brightness profiles in the rest-frame $B$-band, which is significantly affected by the contribution of young stars, we argue this change may be related to the well known cosmological evolution in global SFR between $z$$\sim$1 and $z$$\sim$0. Medians and standard deviations for the distributions represented in Fig. \ref{figMuz} are given in Table \ref{tblfigMuz}.

If we apply here restrictions both in \mb (\mbrange) and \mstar (\mstarrange) to select the objects in all redshift bins the results do not change perceptively either, as in previous cases. The difference in median values of \mubreak between z$\sim$1 and z$\sim$0 is -3.5$\pm$0.3, only slightly larger than for unrestricted samples. This corresponds to a decrease in intensity by a factor 25.1$\pm$8.0. In this case the samples reduce to 11, 11, 39, and 19 galaxies in the reshift bins termed as ``local'', ``low'', ``mid'' and ``high''.

\subsubsection{\rbreak / \hone Evolution.}

\begin{figure}
\epsscale{0.7}
\plotone{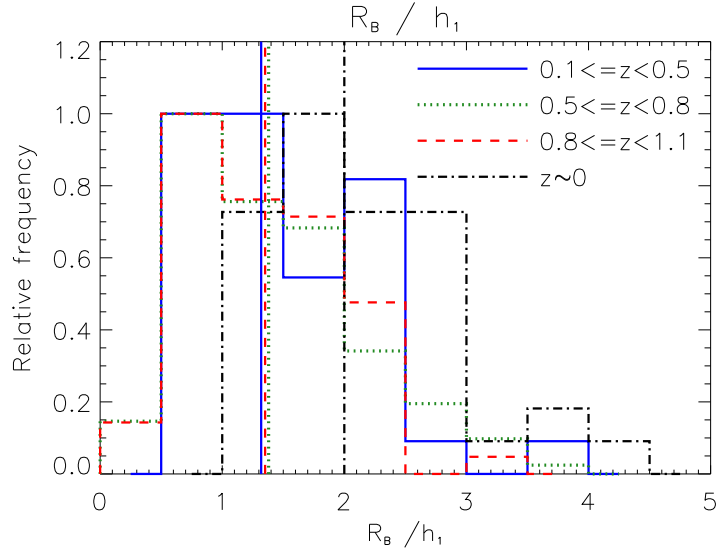}
\caption{Histogram of the ratio \rbreak / \hone for objects of the T-sample, in 3 different redshift bins, and the results in \citet{PT06} at z$\sim$0 ($g'$-band results). The vertical lines mark the median values of \rbreak / \hone for each distribution.}\label{figTrh1z}
\end{figure}

We also present results on \rbreak, relative to the scale length of the inner exponential, \hone, in our truncated objects (T-sample). In Fig. \ref{figTrh1z} are shown histograms of the distribution of this parameter in the 3 explored redshift bins. Also represented is the median value of the ratio for the local sample in PT06, \rbreak/\hone(z$\sim$0) = 2.0 (measured in $g'$-band). The statistical parameters of the distributions can be found in Table \ref{tblfigTrh1z}. The most striking feature is the low set of values of the ratio \rbreak/\hone we have measured, compared to local values. The mean, median and standard deviation of the distributions of values in the different redshift bins are given in Table \ref{tblfigTrh1z}. The median values are also shown as vertical lines in Fig.\ref{figTrh1z}. \citet{Perez04} found a median/mean value of 1.8 for \rbreak/\hone at z$\sim$1 (over 6 objects), which is also shorter than the local value, but still larger than those we find. It is probable that the difference in the sizes of the samples is the cause of this difference. With regard to the difference between the values of \rbreak/\hone she obtained and those observed in the Local Universe, she attributed it mainly to two biases: 1) the detection limit on surface brightness would have prevented her from identifying the breaks with larger value of \rbreak/\hone; and 2) the effects of dust. The first reason is evident, but the latter may need clarification. She assumes dust absorption was more important in late-type galaxies at intermediate redshifts, and specially in the inner parts of the galaxies. This would be perceived as the inner parts of disks having their profiles ``flattened'', i.e. with larger \hone. This is what would make the quotient \rbreak/\hone smaller at higher redshift, relative to the Local Universe. Again in Fig. \ref{figTrh1z}, it is also interesting how the distributions of the ratio \rbreak/\hone are quite similar amongst the sub-samples between z$\sim$0.3 and z$\sim$1.

It is important to note that the \rbreak/\hone evolution presented here is measured in relation to the observed (uncorrected for dust) \rbreak/\hone of the local galaxies; consequently, if the dust opacity were not to change with redshift, the observed evolution presented here would reflect an intrinsic evolution in the profile morphology of the objects. However, it is likely that the opacity of the galaxies changes with redshift. At a fixed inclination, bulge-to-total flux ratio, and rest-frame wavelength, the degree of attenuation and the increase in the observed scale length due to dust can be parameterized by the change in the central face-on optical depth. The optical depth is a very uncertain quantity (even in the nearby Universe), and this makes a detailed evaluation of the effect of dust to go beyond the scope of this work. Consequently, we have not made any attempt to correct our results for the effect of opacity. Nevertheless, in order to provide a crude estimate of how a significant increase in opacity could affect our results, we have performed the following exercise: let us assume a mean inclination of 30 degrees and an increase in the total central face-on optical depth in the $B$-band from 4 (present-day galaxies) to 8 (high-z galaxies). This change implies a transition from an intermediate to a moderately optically thick case. In this case, for a disk-like galaxy observed in the $B$-band rest-frame, the attenuation increases by 0.25 mag \citep[][, their Fig. 3 and Table 4]{Tuffs04} and the scale length increases by 12\% \citep{Mollenhoff06}. If we account for these numbers, the galaxies in our high-z sample would be intrinsically brighter by 25\% and their scale length intrinsically smaller by 12\%. In this sense, if we assume that the break position is not affected by the dust content, the observed (uncorrected for dust increase) \rbreak/\hone evolution presented in this paper would be an upper limit to the actual evolution. It is important to stress however, that a $\sim$15\% bias in \rbreak/\hone due to dust is far from explaining the amount of evolution observed here (\rbreak/\hone increasing by a $\sim$50\% between z$\sim$1 and z$\sim$0, as shown in Fig. 10 and Table 3). Finally, there is no clear evidence that the mean optical depth in dust was in fact so much higher in previous epochs, and if the opacity were smaller in the past, then the situation would be reversed, with our current estimate of the size evolution being a lower limit.

It could be argued that the low values of \rbreak/\hone found at intermediate redshifts could be related to the presence of stellar bars, as objects which present them tend to have lower values of this ratio. In fact, in Pohlen \& Trujillo (2006) it was reported that in galaxies with bars (Type II.o-OLR) \rbreak/\hone is, as measured in $r'$-band, around 1.7 compared to 2.5 in galaxies without bars Type II-CT. However, using the largest up-to-date sample of face-on galaxies (0.2$<$z$<$0.84) from the COSMOS 2-square degree field, \citet{Sheth08} find that the fraction of barred spiral galaxies declines with redshift. For galaxies with \mstar$>$$10^{10}$\msun the fraction of bars drops from 65\% in the local universe to 20\% at z$\sim$84. It must also be taken into account that the Type II.o-OLR objects are included in our ``local'' sample, as, due to insufficient angular resolution, it is not possible to discriminate with the same accuracy in the higher redshift samples, as it is possible to attain with the local objects, whether the ``truncations'' are of this Type (II.o-OLR) or not. These facts together imply that the systematical decline in the \rbreak/\hone ratio we see with redshift is unlikely to be due to a bar fraction increase.

Another point which is interesting to discuss with regard to this problem is related to the apparent similarity of many of the high-z galaxies under study in this work with the irregular galaxies in the Local Universe. Moreover, the values of \rbreak/\hone and \hone are similar to those derived in \citet{HunterElmegreen06} for a sample of irregular galaxies. However, it is unlikely that the present-day irregular galaxies could be the final stage of the galaxies in our sample. Our high-z galaxies are much more massive than these local irregular galaxies. An interesting feature, however, shared by these two families of galaxies is a large value of their specific star formation rate sSFR \citep[see for example the comparison between the sSFR fo $10^{8}$\msun galaxies at low redshift and the $10^{10}$ \msun galaxies at z$\sim$0.85 in Fig. 2 of][]{Bauer05} and the fact that their star formation is well spread over the galaxy disk. This is probably the main reason for their large inner scale-length. 

\begin{figure}
\epsscale{1}
\plottwo{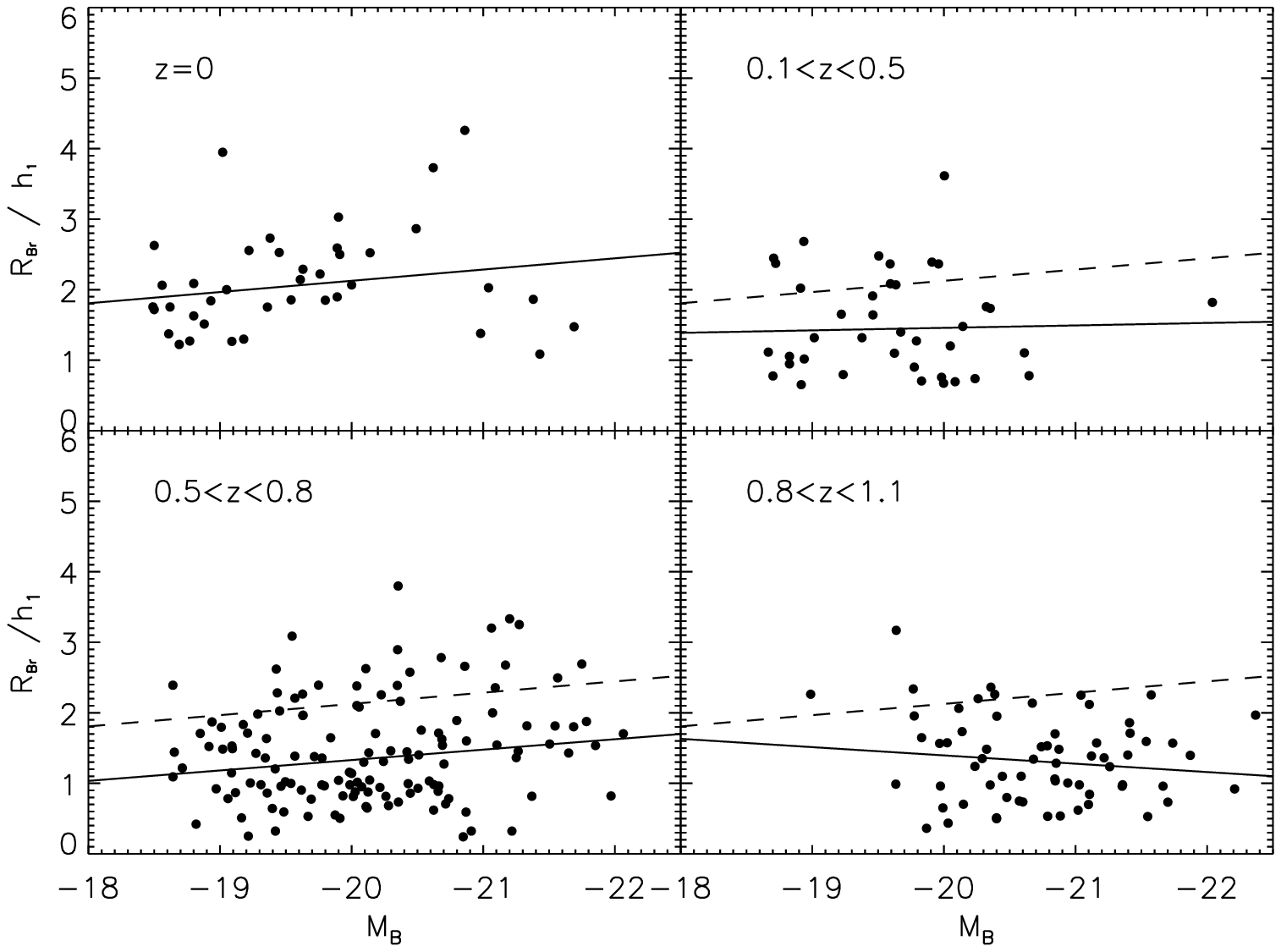}{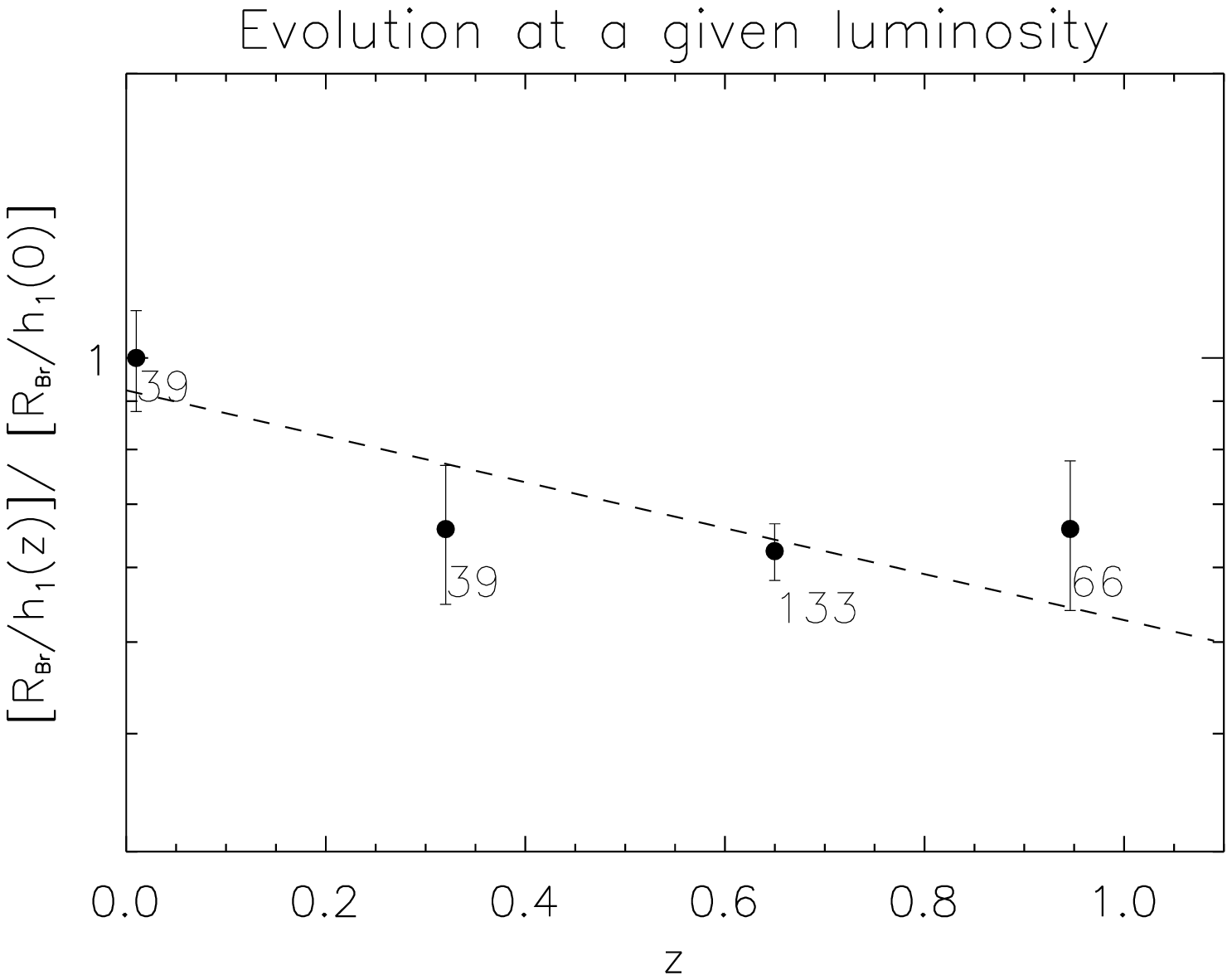}
\caption{Left: \rbreak / \hone for Type II galaxies as a function of absolute $B$-band magnitude \mb, for 4 ranges of redshift. Local data from PT06 ($g'$-band results). Right: \rbreak / \hone at \mb=-20 mag, given by the best fit lines in the left panels, relative to z=0, for each redshift range. The numbers acompanying each point in the right panel give the population of objects which they represent.\label{figTrh1mag}}
\end{figure}

With the purpose of shedding some light on this puzzling phenomenon, we have also probed how \rbreak/\hone relates to global properties of the galaxies. In Fig. \ref{figTrh1mag} (left) we present this ratio against \mb for the T-sample, in analogous way to Fig. \ref{figTrmag}. Again, we have divided the objects in 3 redshift bins, and the local data are from PT06 ($g'$-band results). In the low redshift bin (0.1$<z\lesssim$0.5) our derived relation depends less on luminosity than that reported by PT06. In the ``mid'' redshift bin, though, both relations show as parallel. In the ``high'' range the relation reverses, in the sense that \rbreak/\hone decreases slightly with the $B$-band luminosity of the galaxies, in contrast to the sense in the Local Universe. This feature has not been previously reported and we must try to explain it. First, in the ``high'' redshift bin there is a lack of low luminosity galaxies which might make the slope of the best fit line tilt downwards. But, if this were a real phenomenon (and we will give more evidence below supporting this), and not an observational effect, what would it mean? Assuming an exponential profile, the ratio \rbreak / \hone is proportional to the difference between the surface brightness of the inner exponential measured at the break and at the center: \rbreak/\hone $\propto$ ( \mubreak - $\mu$(r=0) ). So what we see is that, at z$\sim$1 the more luminous objects had a profile which changed less in surface brightness, in absolute terms, from center to the break, than the less luminous objects (i.e. the surface brightness profile is flatter for the more luminous galaxies). Flat profiles, in these rest-frame wavelengths, imply that the star formation was taking place more homogeneously across the whole galaxy disk (a good example is UDF3372 shown in Fig. \ref{figU2G}). 

In the right panel of Fig. \ref{figTrh1mag} we can see the ratio \rbreak/\hone relative to the local value (PT06), at \mb=-20 mag, for every redshift bin, taken from the linear fits given in the left panel. We see how there is little evolution in the ratio from $z$$\sim$0.3 to z$\sim$1. The ratio \rbreak/\hone varies between 0.62$\pm$0.04 and 0.66$\pm$0.11 of the local value at these redshifts, although the error bars are large to derive any firm conclusion. The points in the panel give the impression that there is a ``gap'' between the values at z$\sim$0 and those at intermediate redshifts, perhaps raising concerns about some kind of difference between samples, other than genuine evolutionary changes in this structural parameter, or variations in the methods employed to study the samples. We must say first that, if we take into account the error bars, the significance of this ``gap'' is significantly reduced. We also want to stress that the analysis methodology employed here is analogous to that used in studying the local objects. But we will come over this point below, where the relation \rbreak-\mstar is broached.

\begin{figure}
\epsscale{1}
\plottwo{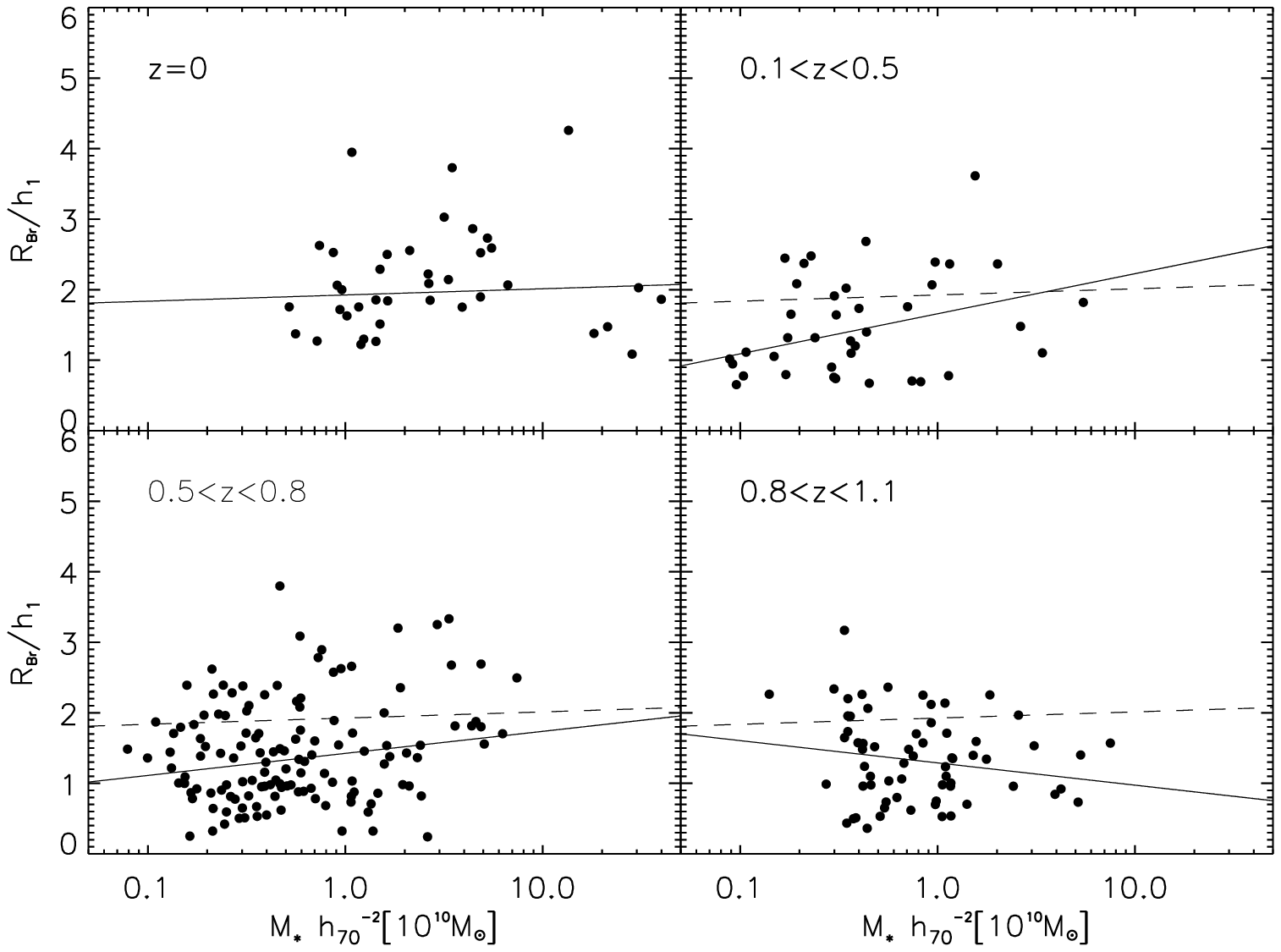}{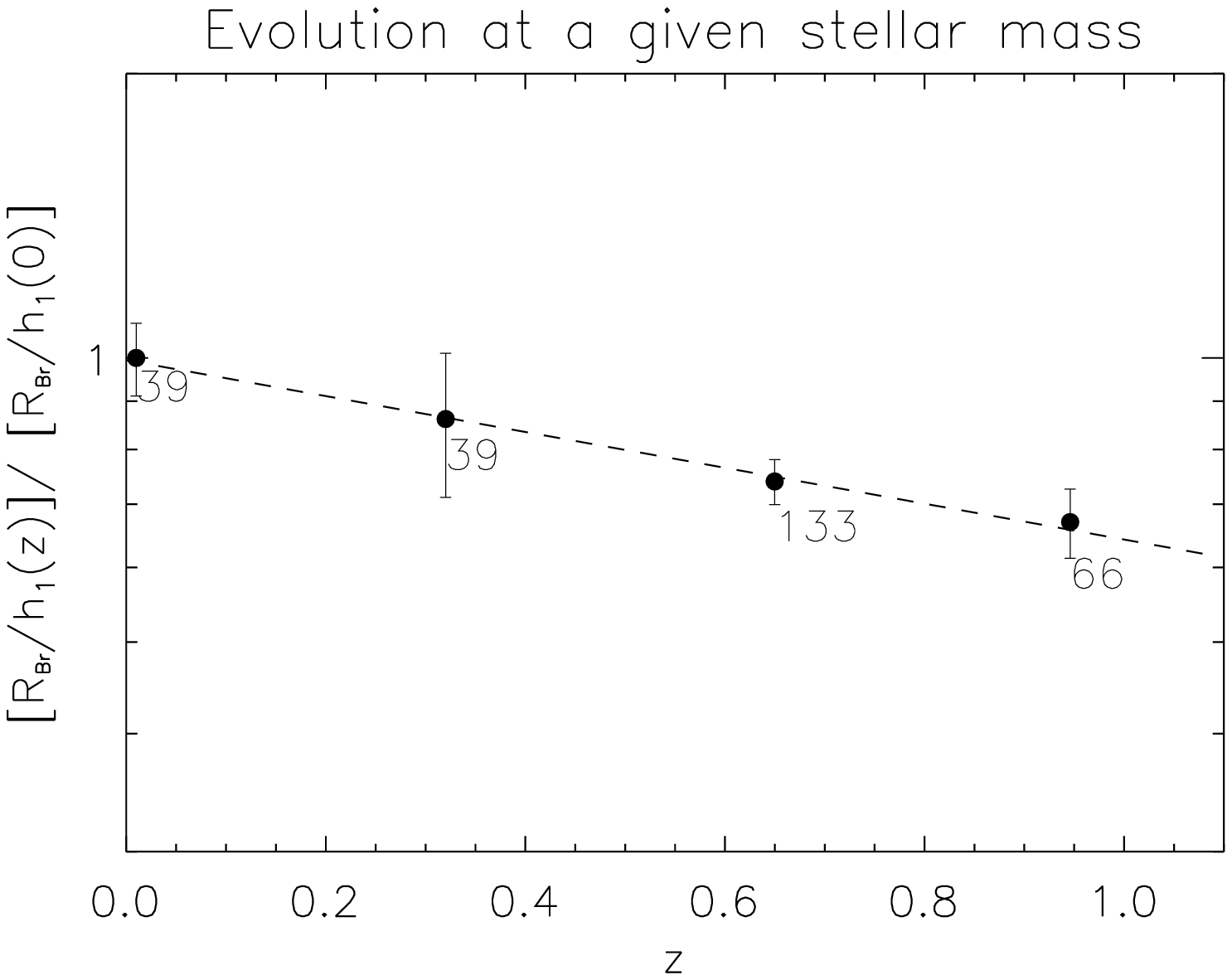}
\caption{Left: \rbreak / \hone of ``truncated'' galaxies as a function of stellar mass \mstar, for 4 ranges of redshift. Local data are from PT06 ($g'$-band results). Right: \rbreak / \hone at \mstar=$10^{10}$ \msun, given by the best fit lines in the left panel, relative to z=0, for each redshift range. The numbers acompanying each point in the right panel give the population of objects which they represent.\label{figTrh1mass}}
\end{figure}

\begin{figure}
\epsscale{1}
\plottwo{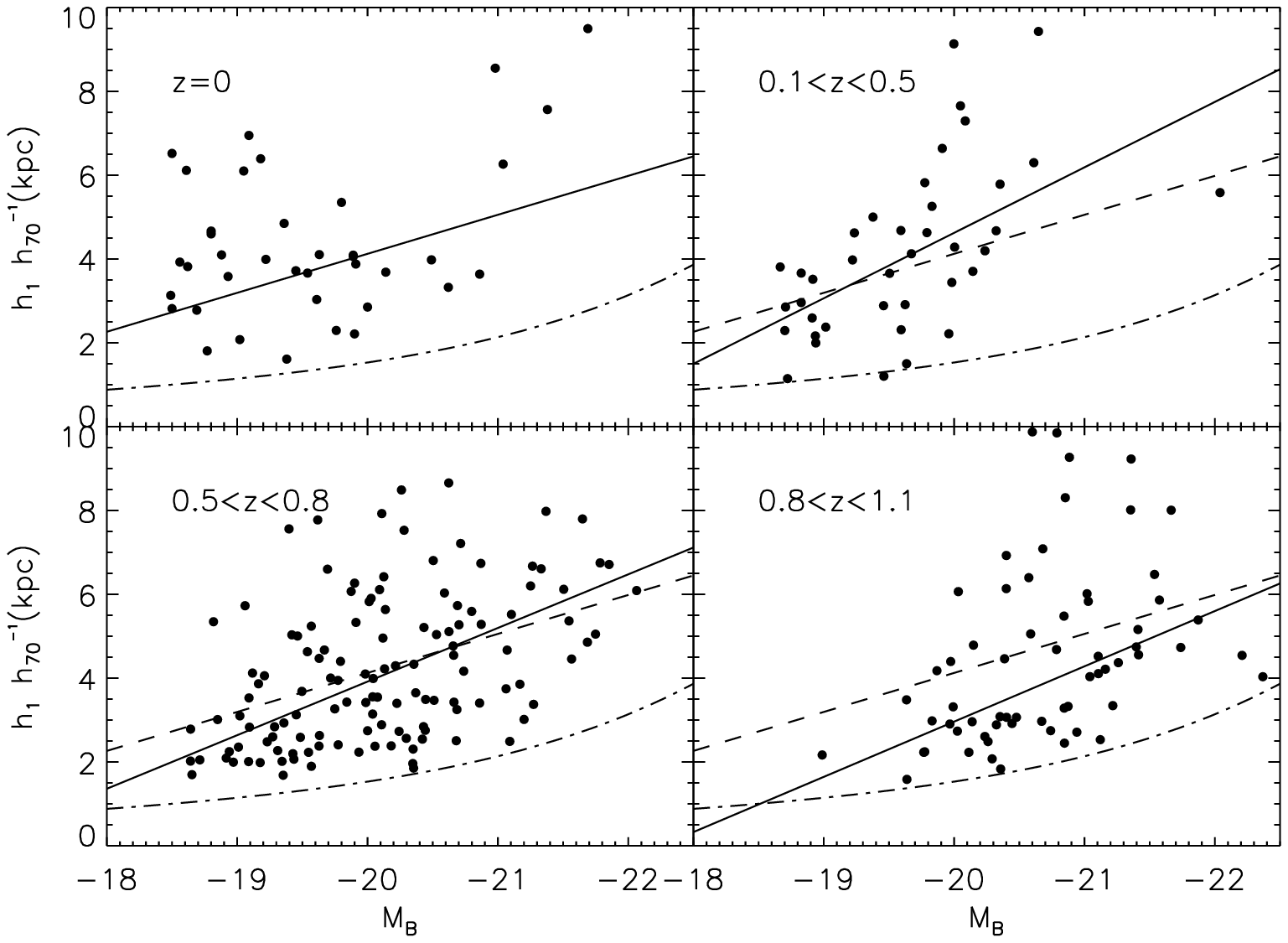}{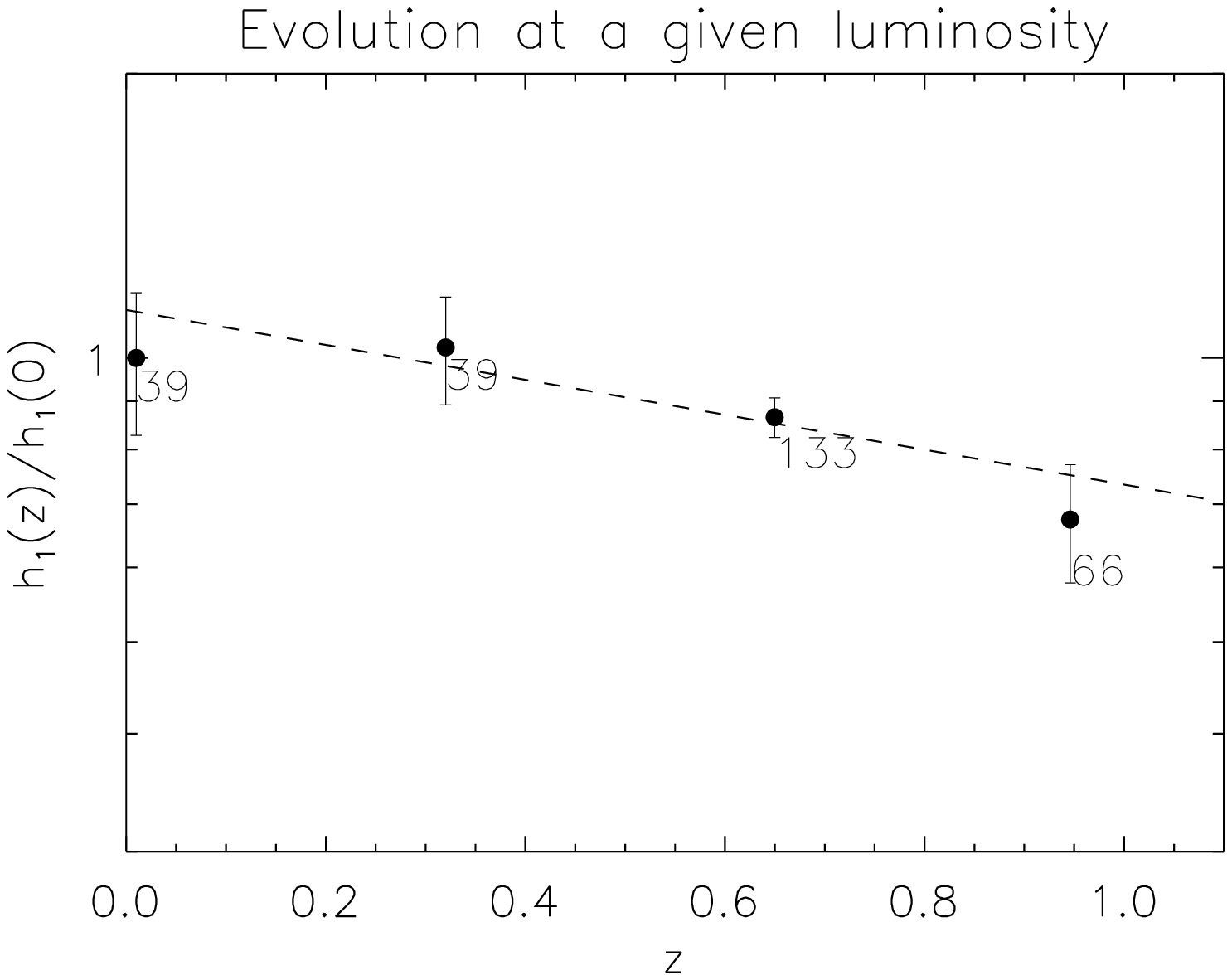}
\caption{Left: \hone for Type II galaxies as a function of absolute $B$-band magnitude \mb, for different ranges of redshift. Local data from PT06 ($g'$-band results). Right: \hone at \mb=-20 mag, given by the best fit lines in the left panels, relative to z=0, for each redshift range. Ordinates are in logarithmic scale. A linear fit to log(\rbreak(z)/\rbreak(0)) - redshift gives a growth by a factor 1.4$\pm$0.5 in \hone between z=1 and z=0 at \mb=-20 mag, a result compatible with no evolution. The slash-dotted line is the relation between size h (equivalent exponential scale length deduced from the S\'ersic effective radius in r' band) and absolute magnitude found by \citet{Shen03} in local late-type (S\'ersic n$<$2.5) objects using the SDSS. Note how, as expected, \hone is larger than h for all the galaxies. The numbers acompanying each point in the right panel give the population of objects which they represent.\label{figh1mag}}
\end{figure}

\begin{figure}
\epsscale{1}
\plottwo{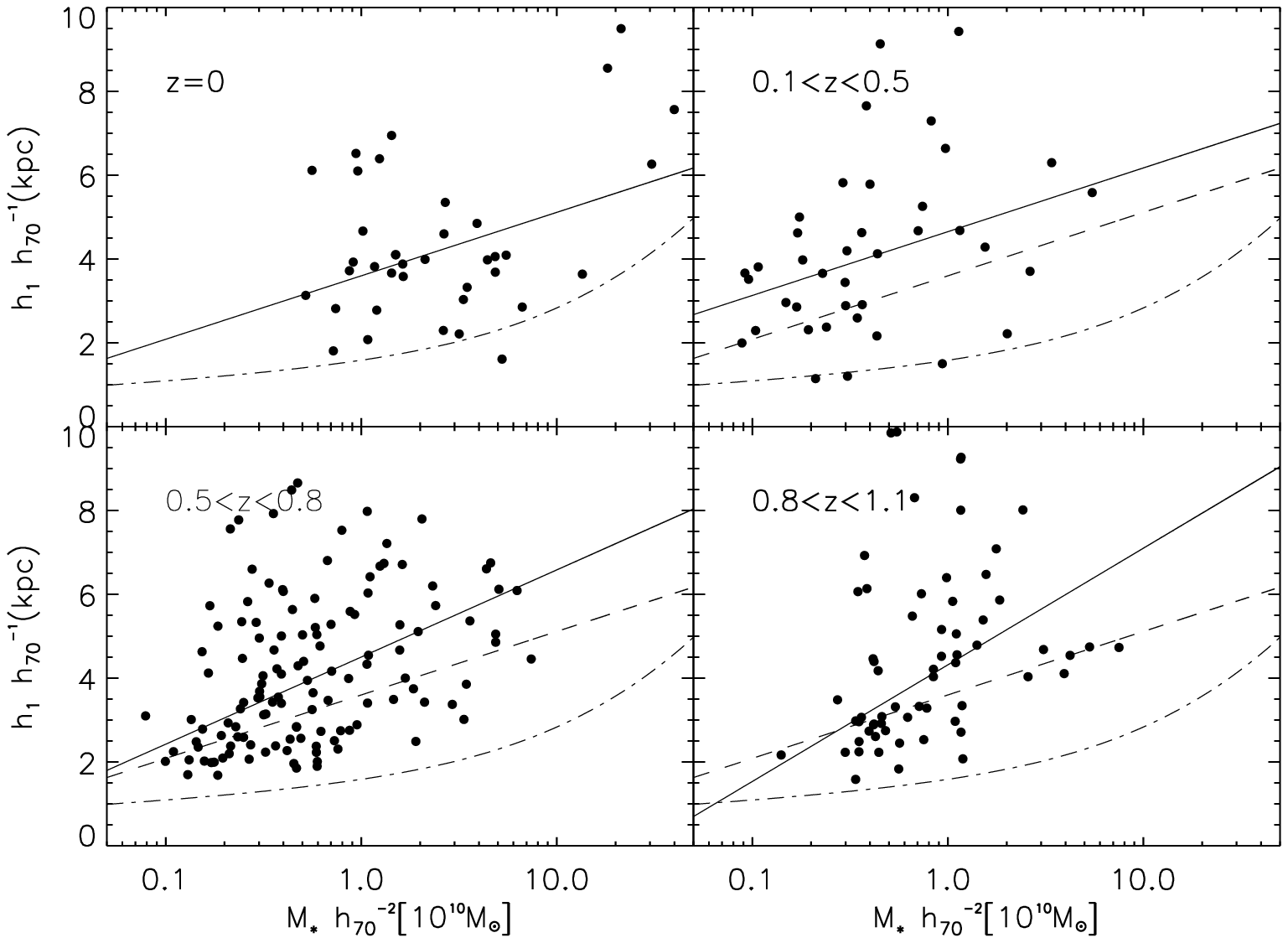}{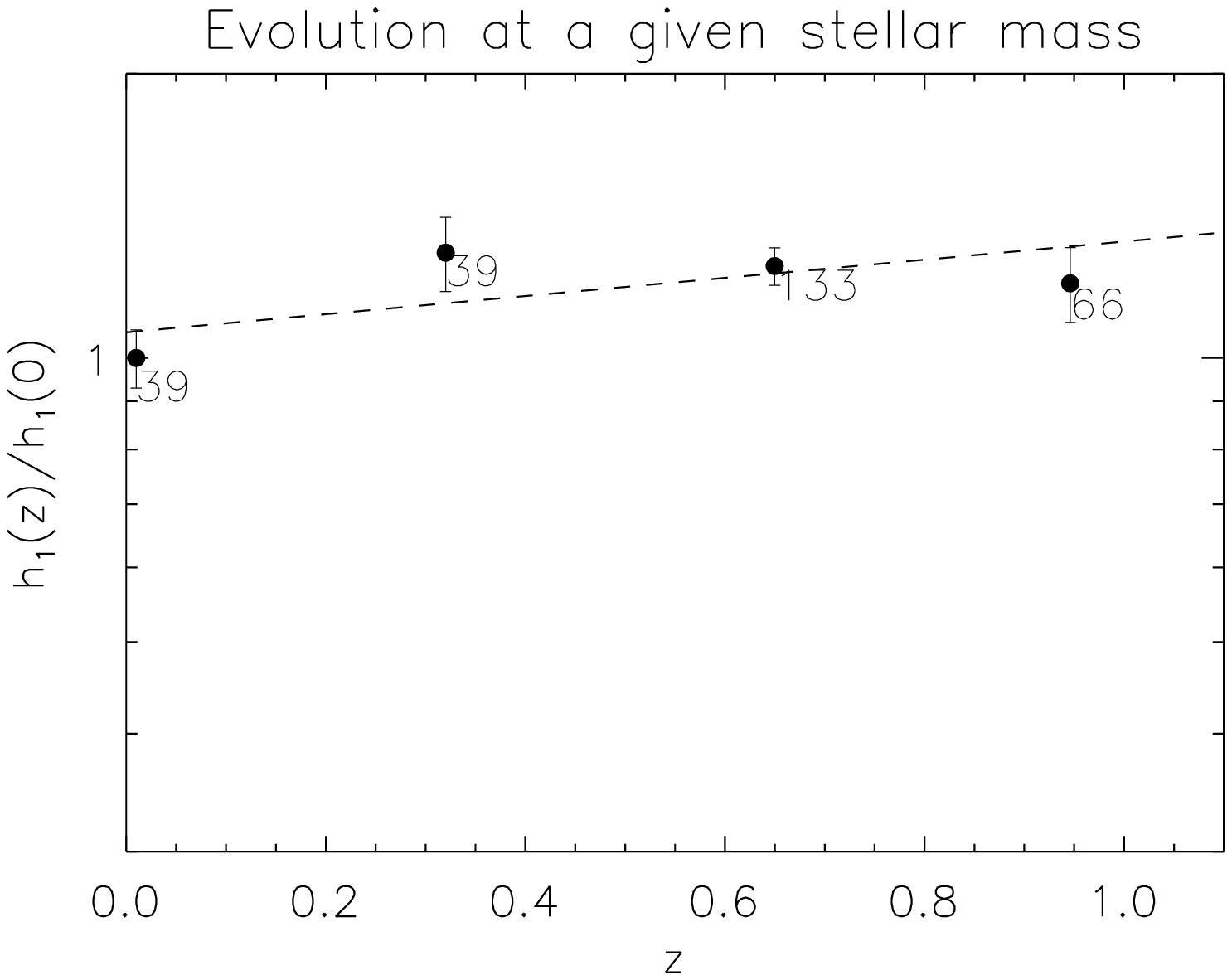}
\caption{Left: \hone of Type II galaxies as a function of stellar mass \mstar, for 4 ranges of redshift. The data for z=0 is from PT06 ($g'$-band results). Right: \hone at \mstar=10 \msun, given by the best fit lines in panel on the left, relative to z=0, for each redshift range. Ordinates are in logarithmic scale. A linear fit to log(\rbreak(z)/\rbreak(0)) gives a shrink by a factor 0.8$\pm$0.1 in \hone between z=1 and z=0 for a same \mstar. The slash-dotted line is the relation between size (equivalent exponential scale length) and stellar mass found by \citet{Shen03} for local, late-type (S\'ersic n$<$2.5) objects in the SDSS (taken from Fig. 11 in their paper). Note how \hone is larger than h for all the galaxies. The numbers acompanying each point in the right panel give the population of objects which they represent.\label{figh1mass}}
\end{figure}

We have also tested how the \rbreak/\hone for the T-sample relates to the stellar mass of the galaxies, and this we can see in Fig. \ref{figTrh1mass}. In the left panel, we see the given ratio against the stellar mass for the different redshift bins (local data are from PT06, as obtained in the $g'$-band). Both panels are analogous to those in Fig. \ref{figTrh1mag}, but substituting \mb for \mstar. In this case, the relation between \rbreak/\hone and \mstar in the ``low'' redshift bin is steeper than the relation reported by PT06 at $z$$\sim$0. Note that the local sample covers a smaller range in stellar mass, which could somehow bias the local \rbreak/\hone determination. In the mid redshift bin, the two relations are almost parallel, and in the high redshift range, the slope changes from positive to negative, as happened in the \rbreak/\hone vs. \mb in the same range of redshifts. It is important to note that the dispersions are quite large, and so these slopes must be taken with caution. But it is interesting to note again an anticorrelation between \rbreak/\hone and \mstar at $z$$\sim$1. As we said above for the more luminous objects, it seems that more massive galaxies at high redshift have an inner disk whose surface brightness decreases less between the central part and the break, compared to the less massive ones. 

In the right panel of the Figure \ref{figTrh1mass} we show the values of the best fits to the distributions of \rbreak/\hone at \mstar = $10^{10}$ \msun, against redshift. This time the decrease in \rbreak/\hone is progressive, and we find that the ratio has increased by a factor 1.6$\pm$0.3 between $z$$\sim$1 and $z$=0, from the linear fit.  

A point of particular interest is that the points in the right panel of Fig. \ref{figTrh1mass} appear better aligned than in the right panel of Fig. \ref{figTrh1mag}. This could mean that the apparent discontinuity in the \rbreak-\mb relation commented above may be, at least in part, caused by the significant evolution in $B$-band luminosity in the same redshift range. But we would not like to push this interpretation, as the error bars are large enough to be compatible with a more progressive evolution of \rbreak/\hone at a given \mb with redshift.

Also for the \rbreak/\hone parameter have we tested its relations with \mb and \mstar when the corresponding ranges to select the objects are fixed, as in previous subsections. In the case for a fixed \mb (selection range: \mbrange), the ratio \rbreak/\hone at intermediate redshifts varies between 0.50$\pm$0.08 and 0.59$\pm$0.17 of the local value. This is a slightly larger difference with local values than for the unrestricted samples. This does not affect in a perceptible way our discussion on this parameter. For the relation between \rbreak/\hone and \mstar (selection range: \mstarrange), the result remains just unchanged with respect to the unrestricted case. This is, there is an increase in the ratio also by a factor 1.5$\pm$0.3 between $z$$\sim$1 and $z$=0, from the linear fit.

\subsubsection{Scale length of the disk inner to the Break: \hone.}

In a further attempt towards clarifying why we find a distribution of \rbreak/\hone with lower mean and median values at higher redshift than in the Local Universe, we have also probed the relation between \hone and \mb and \mstar for the ``truncated'' galaxies under study.

In Fig. \ref{figh1mag} we present \hone as a function of \mb for galaxies of the T-sample, again for different redshift bins, as in previous similar figures (\ref{figTrmag} and \ref{figTrh1mag}). In the left panel we have also included the relation between ``size'' (i.e. scale length derived from the effective radius as explained below) and absolute magnitude found by \citet{Shen03} for local, S\'ersic $n<$2.5 galaxies in the SDSS, represented by their best fit curve (slash-dotted line). \citet{Shen03} published the relation between the S\'ersic effective radius of the galaxies (i.e. the radius that encloses half of the light of the S\'ersic model that best fits to the distribution of light of the object), as measured in $r'$ band and absolute magnitude (see Fig. 6 in their paper). Our profiles are retrieved in approximately rest-frame $B$-band ($\sim$ $g'$ band), but we would not expect dramatic differences in the size estimates of disks  between the two bands (either from scale lengths or truncation radii; see eg. values of these parameters in both bands in Table in PT06). Assuming an $n$=1 S\'ersic profile (exponential profile), we converted the effective radii ($R_{S,eff}$) to scale length of the exponential profile using the expression $h$ = $R_{S,eff}$ / 1.676. As their cut-off for classification as a late-type galaxy was as broad as $n<$2.5 (the same as ours), this implies that most of their objects do not obey the exponential law, and we should term this $h$ as the equivalent scale length. We still show their curve for reference. First, we see how the values of \hone in the PT06 sample fall above the \citet{Shen03} curve (\hone values from PT06 were obtained in the $g'$-band). This result is expected since fitting the whole profile of a truncated galaxy will produce smaller values of scale length than for the inner part. Our results are divided amongst the other 3 sub-panels as a function of redshift, and we see how the best fit lines to the distributions have much steeper slopes than for the local sample. Our galaxies cover a wider range in \hone relative to local ones, reaching larger limits, especially for the most luminous objects. We see in the right panel the \hone values from the best fits to the distributions at \mb=-20 mag, as a function of redshift. It can be seen how there is a moderate decrease in \hone with redshift. 

In Fig. \ref{figh1mass} (left) we show the relation between \hone and the stellar mass for the galaxies of the T-sample, at different redshifts. The local data, as before, are from PT06 (results obtained in $g'$-band). The slash-dotted line represents the relation between equivalent scale length, $h$ = $R_{S,eff}$ / 1.676, and stellar mass given in \citet{Shen03} (taken from their Fig. 10). That is a best fit to the relation they find for late-type ($n<$2.5) galaxies in SDSS, in the $z'$ band. Again, we would not expect dramatic changes in size parameters of disks between $B$ and $z'$ band, and this is the best suited result for comparison they give. It is evident that as in the \hone - \mb relation, the values of \hone fall systematically above the \citet{Shen03} relation. The reason for this is the same as the given previously: the inner scale lengths in ``truncated'' galaxies are systematically larger than the scale-length of the whole disk. We can see how our distributions of \hone are slightly above (reaching higher values) those for the local sample, and their best fit lines have steeper slopes than in the lowest redshift range. In the right panel of the same Figure we see the best fit value to the \hone distributions at \mstar = $10^{10}$\msun, relative to the local value, against redshift. The values of \hone at intermediate redshifts are larger than the local ones by a factor that goes from 1.29$\pm$0.12 to 1.20$\pm$0.11 between $z\sim$0.3 and $z\sim$0.9. This overall slight increase in \hone, combined with the decrease of \rbreak are responsible for the reported decrease in \rbreak/\hone at higher redshift. The decrease by a factor 1.3 in \rbreak (at a fixed stellar mass), times the increase in \hone by $\sim$1.25 (also at a given stellar mass) gives 1.6, in agreement with the reported decrease between $z\sim$0 and $z\sim$1. From the lack of evolution of \hone at intermediate redshifts seen in the right panel of Fig. 14, and the previous discussions on the \rbreak - \mstar and \rbreak/\hone relations, we can also say something else. The disks have become fainter (in rest-frame $B$-band), while keeping their slopes ($\propto$1/\hone) roughly constant or slightly increasing, and at the same time increasing their \rbreak, as the Universe has grown older (in the surveyed range of redshifts).

We have also probed these relations under restrictions in the selected \mb (\mbrange) and \mstar (\mstarrange), depending on whether \hone is related to luminosity or stellar mass respectively. On one hand, for the \hone - \mb relation, this varies by a factor 1.02$\pm$0.2 between z$\sim$1 and z$\sim$0; i.e. no evolution in the surveyed range of redshifts is retrieved (in the ``unrestricted'' case this factor is 1.4$\pm$0.5). On the other hand, for the \hone - \mstar relation, the result is the same as in the case when no restrictions in \mstar are applied. This is, from the linear fit we obtain a decrease in \hone by a factor 0.8$\pm$0.1 between z$\sim$1 and z$\sim$0 when the mass is fixed. Again we see how using the same ranges of luminosities and stellar masses in every redshift bin does not vary significantly the results.

\subsubsection{Robustness of the reported evolutionary changes.}

The previous sub-sections have shown that both the break position \rbreak and the surface brightness at the break radius \mubreak have evolved with cosmic time. It could be argued, however, that part of this evolution could be due to selection effects, since as shown in Section 4 very faint breaks can be missed if they were present in the galaxies analysed. We have checked whether this is the case by exploring how our surface brightness limit could affect the observed evolution.

According to Sec. 4, the profiles are reliable down to $\mu$$\sim$26 \magarcsq. Being conservative, we estimate that we should be able to identify breaks with high confidence down to \mubreak$_{,limit}$$\sim$25.5 \magarcsq. Using this number as the limiting value we could estimate what would be the rest-frame (i.e. cosmological dimming corrected) surface brightness at the break at which the number of objects classified as of Type II will start to decline due to a selection effect and not as a result of a real evolution. These numbers will be 24.4 \magarcsq ($z\sim$0.3), 23.3 \magarcsq ($z\sim$0.65) and 22.6 \magarcsq ($z\sim$0.95). As we can see in Fig. \ref{figMuz}, these numbers are far away (more than a magnitude in all cases) from the position of the peak in the observed galaxy surface brightness break distribution. This reinforces the idea that most of the observed evolution is real and not caused by this effect.

Another test we have run is to estimate what would be the maximum \rbreak that could be measured at a given absolute magnitude according to the limiting surface brightness proposed above. At the highest redshift bin, $z\sim$1, the ``distance modulus'' is 43 mag ($k$ correction included, derived from the direct comparison of \zz magnitudes to \mb). Again, we use a surface brightness limit of 25.5 \magarcsq. We consider a truncated exponential galaxy with a typical scale length of $h$ = 0.5 arcsecs. With these values, the largest \rbreak that could be measured would be: \rbreak = 11.1 kpc for galaxies with \mb=-20 mag and \rbreak=18.5 kpc for galaxies with \mb = -22 mag. As can be seen these values are again far away from the observational data distribution, indicating that most (if not all) of the observed evolution is real and not caused by our limiting detected surface brightness.

\section{Conclusions}\label{sec6}

In this work we have presented an analysis of surface brightness profiles for a sample of 505 galaxies in the GOODS-South field, making use of HST-ACS imaging from \citet{Giavalisco04}. These galaxies are all classified as disk galaxies, based on the selection of those objects with S\'ersic indexes $n<$2.5 \citep[S\'ersic profile fitting is from][]{Barden05}. We have also used publicly available data on redshift estimates, absolute rest-frame $B$-band magnitudes and stellar masses from \citet{Wolf01,Wolf03}, \citet{Barden05}. We have classified the profiles using the presence of ``breaks'' in the exponential profiles of the disks, and the ratio of inner and outer scale lengths (\hone/\htwo) in Types I (no apparent break), II (``downbending break''; \hone/\htwo$>1$) and III (``upbending break'' ; \hone/\htwo $<$1). This characterization has been performed in bands which, within the explored range of redshifts, track the rest-frame $B$-band.

We have performed simulations on the classification and profile characterization of artificial galaxies, analogous to that performed on real galaxies. With regard to their classification, the worst results are for Type I objects, 50 \% of which we fail to classify correctly at the faintest magnitudes (\zz$>$22.5 mag AB), while for Types II and III the success rates are much higher ($\gtrsim$70-80\%). We have also compared our results on classification and \rbreak estimate to those presented in \citet{TP05} for a set of 32 galaxies in common to our sample and theirs. Their results are based on the \emph{HST}-UDF observations, which are deeper than the GOODS images. Both sets of classifications match in $\sim$70\% of cases. Particularly, in the case of Type II galaxies there were only 2 cases out of 19 ($\sim$10\%) that were classified differently in our work due to insufficient depth in the images. When we compare the \rbreak and \mubreak estimated in the two studies for Type II objects in common, we find no clear bias in any sense. The dispersions are $\sigma_{R}^{err}$=22\% and $\sigma_{\mu}^{err}$ = 43\%, which can be used as estimates of the errors in these structural parameters.

In this work we put special emphasis on the study of ``truncated'' galaxies, i.e. those with Type II profile, with the ``break'' taking place in the outer part of the disk. This subsample, which we call the T-sample, is composed of 238 objects, an order of magnitude larger than any sample in previously published work at intermediate redshifts. We have studied the relation between the radius at which the ``break'' takes place, \rbreak, and absolute $B$-band magnitude and total stellar mass for these objects. We find a clear evolution in \rbreak with redshift, galaxies with the same values of luminosity/stellar mass having shorter \rbreak's than local galaxies \citep[using as reference the results in][]{PT06}. We measure an increase by a factor 1.3$\pm$0.1 in \rbreak between $z\sim$1 and $z\sim$0 at fixed \mstar = $10^{10}$ \msun, and by a factor 2.6$\pm$0.3 at fixed \mb = -20 mag, in the same range of redshifts.

At the same time, there is also clear evidence for a decrease in the surface brightness level \mubreak at which the ``break'' takes place on the profile in the last $\sim$8Gyr. We find that \mubreak (in the rest-frame $B$-band) has decreased by 3.3$\pm$0.2 mag between $z\sim$1 and nowadays. This is equivalent to a decrease in surface intensity by a factor 20.9$\pm$4.2.

Another point of interest is how \rbreak relates to the scale length \hone of the disk inside the ``break''. We show results on the ratio \rbreak/\hone at different redshift ranges up to z$\sim$1. The median values of this ratio are significantly smaller than those found in the Local Universe (1.4$\pm$0.6 at $z\sim$1 versus 2.0$\pm$0.7 at $z\sim$0). We also find that in the highest redshift bin (0.8$<z\leq$1.1) \rbreak/\hone tends to decrease moderately with \mb and \mstar, in contrast to what happens in the Local Universe and the other redshift bins. This means that, at the epoch corresponding to $z\sim$1, more massive objects had somewhat smaller difference in brightness between the central part of the disk and the break radius, than less luminous / massive objects (implying that their surface brightness profiles are flatter).

Finally, we have also probed the relation between \hone and $B$-band luminosity and stellar mass. We see how the distribution of values of \hone is broader, reaching higher values, at intermediate redshifts than in the Local Universe. The mean values of \hone seem to be somewhat larger than local ones, by a factor 1.3 at most, but there is almost no evolution in them (within our statistics) between $z\sim$0.3 and $z\sim$1.

We have also tested the previous relations when the samples are selected within the same ranges of \mb and/or \mstar in every redshift bin. The reported figures do not change significantly under these restrictions, and this further supports the robustness of the results.

We conclude by remarking that our results are consistent with the following picture of disk evolution. We find that in the lapse of time that goes from $z\sim$1 to the present epoch ($\sim$8 Gyr), the radii of the ``break'' in truncated disks, as measured in $\sim$$B$-band, have increased noticeably, while the intensities (on the profile) at which the break takes place have substantially decreased. We interpret the \rbreak, in this particular case, as a measure of the size of the disk where most of the massive star formation is taking place, as the $B$-band is more influenced by the emission from these young stars than redder bands. At a given stellar mass, the scale lengths of the disk in the part inner to the ``break'' were on average somewhat larger in the past, and have remained more or less constant until recently. This phenomenon could be related to the spatial distribution of star formation, which seems to be rather spread over the disks in the images. So disk galaxies had profiles with a flatter brightness distribution in the inner part of the disk, which has grown in extension, while becoming fainter and ``steeper'' over time. This is consistent with at least some versions of the inside-out formation scenario for disks.

\acknowledgments
We are grateful to Marco Barden for kindly providing us with the GEMS morphological analysis catalog. We acknowledge the COMBO-17 collaboration (especially Christian Wolf) for the public provision of a unique database upon which this study is based. We also thank Michael Pohlen for permitting us to use the local sample data for comparison. We must acknowledge the GOODS team for providing such a valuable and easily accesible database from which we could extract our results. We also thank the anonymous referee for insightful and fruitful revision of the manuscript from which its quality has much benefited. This work is based on observations made with the NASA/ESA \emph{Hubble Space Telescope}, which is operated by the Association of Universities for Research in Astronomy, Inc., under NASA contract NAS5-26555. Partial support has been provided by projects AYA2004-08251-C02-01 and AYA2007-67625-C02-01 of the Spanish Ministry of Education and Science, and P3/86 of the Instituto de Astrof\'isica de Canarias.



{\it Facilities:} \facility{HST (ACS)}



\appendix
\section{Testing the Reliability of the characterization of Profiles.}
\label{appendix}

In this Appendix we explain the work done on the profile characterization of synthetic galaxies, relevant to ``breaks''. We have generated 300 2D models of galaxies using GALFIT \citep{Peng02}, which are simplified emulations of real galaxies observed in the range of redshifts we explored. The GALFIT software allows the user to build a model as the sum of a series of analytic 2D components. For each component the user can choose the 2D geometry of the isophotes, as a generalized ellipse (i.e. with ``regular'' appearance, but which can also adopt a ``disky'' or ``boxy'' appearance, depending on the ``c'' parameter), and besides, also the law that governs the radial profile of intensity. The user can choose amongst different laws: exponential, S\'ersic 1/n profiles, gaussians, moffatians, and more. As we are interested in the mid to outer parts of disks, we used only exponential behavior, without introducing bulges or other features. It is possible to reproduce a ``broken'' exponential disk as the composition of 2 concentric exponential disks, with the same isophote geometry, cancelling each disk in the internal or external region, relative to the radial position of the break. Although we produce these ``broken disks'' as the combination of two exponential components, we then define these models as containing a single component; i.e. a disk with a ``break'' in its profile.

All the models synthetized have only one component, with a geometry of the isophotes invariant with radius. We chose an elliptical geometry with c=0, i.e. a common ellipse (neither ``disky'' nor ``boxy''). The axis ratio q for each model was randomly chosen in the range 0.5$<$q$<$1, to mimic variations in orientation of the disks. Three kinds of models, each corresponding to a Type of profile (I, II and III) were generated. For Type I galaxies, the model follows a single exponential law. The other 2 kinds of models have also a single component, but they follow an exponential law with a slope that changes from \hone to \htwo at some radius, the break radius (\rbreak), using the technique described above. Depending on the ratio \hone / \htwo, with this kind of model we represent either a Type II (\hone / \htwo $>$ 1) or a Type III (\hone / \htwo $<$ 1) galaxy. We generated 300 models, 100 of each of these 3 kinds (Types I, II and III). The ranges of \hone, \htwo, \rbreak, \mubreak and total magnitudes from where to extract the parameters to build each type of model are shown in Table \ref{tblSim}. The ranges bracket the whole space in which the parameters vary in our sample of real galaxies, at all redshifts. Some further restrictions are applied to discard certain combinations of parameters, which would produce models which are classified as wildly unreal or undetectable by SExtractor: a) the ratio \rbreak / \hone is limited to be $<$ 6 ; b) the magnitude in the \zz band is within 21$\leq$\zz$\leq$24 mag ; and c) the central intensity must be above 3 $\sigma_{sky}$ of the simulated image.

The 300 created objects were placed in a synthetic, noisy \zz band image with the same pixel scale, zero point and noise level as those of the GOODS data. Crowding effects were not simulated as we have discarded in our analysis those real galaxies which are too near to other objects. SExtractor was run on the synthetic image, using the same detection parameters as those we used for real data, to produce a photometric catalog and segmentation map. These were then used to extract the radial profiles of the objects, and we analysed them with the same software we used for studying the real objects. Following the same criteria we used for real galaxies, one of us visually classified each model as being of Type I, II or III. If he judged there was any break in the profile, he pinpointed its radial and surface brightness coordinates in the \mur profile. The characterization test was ``blind'', in the sense that the classifier did not know a priori the parameters that defined the models he was facing, and all models were ``shuffled'', so as to avoid bias. The results of this simulation and the constraints they imply on the reliability of our results on analysis of real objects have been explained in section \ref{sec3}.




\clearpage



\clearpage
\input{stub.tab1}
\clearpage

\begin{table}
\begin{center}
\caption{Characterization of histograms in Figure \ref{figMuz}.}\label{tblfigMuz}
\begin{tabular}{r|rrr}
\tableline\tableline
                     median & $\sigma$ \\
\tableline
z$\sim$0              23.8\tablenotemark{\dag}   &  0.8	\\
$0.1 < z \leq 0.5$    22.3  &  0.98	\\
$0.5 < z \leq 0.8$    21.5  &  0.81	\\
$0.8 < z \leq 1.1$    20.6  &  0.79	\\
\tableline
\end{tabular}
\tablenotetext{\dag}{g' band}
\tablecomments{Surface brightness at the break (\mubreak) in truncated galaxies.}
\end{center}
\end{table}

\begin{table}
\begin{center}
\caption{Characterization of histograms in Figure \ref{figTrh1z}.}\label{tblfigTrh1z}
\begin{tabular}{r|rrr}
\tableline\tableline
                    &  mean & median & $\sigma$ \\
\tableline
z$\sim$0            & 2.12  &  2.00  &  0.73    \\
$0.1 < z \leq 0.5$  & 1.52  &  1.32  &  0.71    \\
$0.5 < z \leq 0.8$  & 1.45  &  1.38  &  0.73    \\  
$0.8 < z \leq 1.1$  & 1.35  &  1.35  &  0.60    \\  
\tableline
\end{tabular}
\tablecomments{\rbreak/\hone ratio in truncated galaxies.}
\end{center}
\end{table}

\begin{table}
\begin{center}
\caption{Simulation Parameters.}\label{tblSim}
\begin{tabular}{c|rrrrrr}
\tableline\tableline
           & \hone       &  \htwo      &  \rbreak     &  \mubreak       &  Total Flux    \\
Type (Number)  & arcsec        &   arcsec      &  arcsec      & \magarcsq &   magnitudes   \\
\tableline
I (100)         & (0.1'', 0.7'') & \nodata       &  \nodata     &  \nodata         & (21, 24)        \\
II (100)        & (0.1'', 1.7'') & (0.1'', 0.5'')\tablenotemark{a} & (0.2'', 1.7'')\tablenotemark{c} &    (22, 26)      & (21, 24)        \\
III (100)       & (0.1'', 0.4'') & (0.1'', 0.7'')\tablenotemark{b} & (0.2'', 1.2'')\tablenotemark{c} &    (23, 26)      & (21, 24)        \\
\tableline
\end{tabular}
\tablenotetext{a}{\hone$>$\htwo}
\tablenotetext{b}{\hone$<$\htwo}
\tablenotetext{c}{\rbreak/\hone $<$ 6}
\tablecomments{Parameters for the creation of artificial galaxies \citep[using GALFIT][]{Peng02}, which enter our simulation on analysis of surface brightness profiles. The models were selected to have central surface brightness $\mu(r=0)$ above 3$\sigma_{sky}$, and only models with \rbreak / \hone $<$ 6, when appliable, are allowed.}
\end{center}
\end{table}

\begin{table}
\begin{center}
\caption{Results on Classification of Profiles.}\label{tblClass}
\begin{tabular}{r|rrrrrr}
\tableline\tableline
                    &  I &  II  & III &  Raw Total\tablenotemark{a}  & Discarded  & Net Total\tablenotemark{b}\\
\tableline
$0.1 < z \leq 0.5$  &  17 (25$\pm$6\%) &   40 (59$\pm$9\%) &   10 (15$\pm$5\%) &    81   &     14     &  67       \\
$0.5 < z \leq 0.8$  &  77 (33$\pm$2\%) &  135 (58$\pm$5\%) &   22 (9$\pm$2\%)  &   273   &     39     & 234       \\
$0.8 < z \leq 1.1$  &  52 (39$\pm$5\%) &   67 (50$\pm$6\%) &   15 (11$\pm$3\%) &   151   &     17     & 134       \\
\tableline
Sum                 & 146          &  242          &   47          &   505   &     70     & 435       \\
\tableline
\end{tabular}
\tablenotetext{a}{Number of objects surveyed at each redshift range.}
\tablenotetext{b}{Number of objects analysed at each redshift range, after rejection of those not well suited to our study.}
\end{center}
\end{table}

\end{document}

%% file: stub.tab1
\begin{deluxetable}{rrrrrrrrr}
\tabletypesize{\scriptsize}
\tablecolumns{9}
\tablewidth{0pc}
\tablecaption{Results on the classification and characterization of Surface Brightness Profiles.}
\tablehead{
\colhead{GEMS} & \colhead{RA} & \colhead{DEC} & \colhead{z} & \colhead{Type} & \colhead{$R_{B}$} & \colhead{$\mu_{B}$} & \colhead{$h_{1}$}\tablenotemark{a} & \colhead{Comment}\\ & & & & & \colhead{kpc} & \colhead{\magarcsq} &\colhead{kpc} &  \\
}
\startdata
GEMSz033300.19m275140.4 & 03:33:00.21 & -27:51:40.40 & 0.15 & II & 15.88$\pm$0.88 & 24.74$\pm$0.18 & 6.64 & \\
GEMSz033302.71m275140.1 & 03:33:02.72 & -27:51:40.16 & 0.45 & II & 5.07$\pm$0.70 & 23.44$\pm$0.13 & 7.29 & \\
GEMSz033256.65m275316.5 & 03:32:56.66 & -27:53:16.50 & 0.35 & III & 4.26$\pm$0.13 & 23.58$\pm$0.14 & 0.77 & \\
GEMSz033249.35m275345.9 & 03:32:49.35 & -27:53:45.95 & 0.40 & II & 3.71$\pm$0.15 & 22.65$\pm$0.20 & 5.26 & \\
GEMSz033253.89m275354.0 & 03:32:53.90 & -27:53:54.02 & 0.15 & II & 11.07$\pm$0.63 & 24.00$\pm$0.23 & 4.68 & \\
GEMSz033249.67m274705.6 & 03:32:49.68 & -27:47:05.35 & 0.47 & I & \nodata & \nodata & 4.95 & \\
GEMSz033255.01m274706.0 & 03:32:55.02 & -27:47:06.01 & 0.50 & II & 3.20$\pm$0.31 & 23.03$\pm$0.27 & 2.91 & \\
GEMSz033255.00m274953.2 & 03:32:54.98 & -27:49:53.16 & 0.41 & \nodata & \nodata & \nodata & \nodata & Irr/Merger\\
GEMSz033250.76m274525.8 & 03:32:50.75 & -27:45:25.89 & 0.29 & \nodata & \nodata & \nodata & \nodata & Edge\\
GEMSz033245.02m275439.5 & 03:32:45.02 & -27:54:39.58 & 0.47 & II & 10.16$\pm$1.05 & 23.47$\pm$0.36 & 5.58 & \\
GEMSz033231.95m275457.2 & 03:32:31.95 & -27:54:57.37 & 0.50 & I & \nodata & \nodata & 3.08 & \\
GEMSz033235.07m275504.9 & 03:32:35.07 & -27:55:05.00 & 0.33 & III+II & 3.85$\pm$0.22 & 24.43$\pm$0.09 & 2.22 & \\
GEMSz033239.38m275507.1 & 03:32:39.39 & -27:55:07.09 & 0.14 & II+III & 3.13$\pm$0.07 & 22.26$\pm$0.05 & 2.37 & \\
GEMSz033239.69m275507.1 & 03:32:39.70 & -27:55:07.17 & 0.35 & I & \nodata & \nodata & 1.56 & \\
GEMSz033245.94m275546.4 & 03:32:45.95 & -27:55:46.47 & 0.23 & I & \nodata & \nodata & 1.82 & \\
\enddata
\tablenotetext{a}{The median value of the Pearson's R parameter for the linear fit from where \hone is derived is \~{r}$_{1}$ = 0.993}
\tablecomments{The name given for the objects is from GEMS \citep{Rix04}. Coordinates are those retrieved by our software on GOODS-HST/ACS images (J2000.0). Redshifts are from COMBO17 \citep{Wolf01,Wolf03}. Stellar masses and B band absolute magnitudes are extracted from \citet{Barden05}. The whole list can be consulted in the on-line edition.\label{tblResults}}
\end{deluxetable}